Politechnika Wrocławska

**Wydział Informatyki i Zarządzania**
kierunek studiów: Informatyka
specjalność: Multimedialne Systemy Informacyjne

Praca dyplomowa - magisterska

# Analysis of Social Group Dynamics

Stanisław Saganowski

short abstract:
In this thesis the method for social group evolution discovery, called GED, is analyzed. Especially, GED method is compared with other methods tracking changes in groups over time with focus on accuracy, computational cost, ease of implementation and flexibility of the methods. The methods are evaluated on overlapping and disjoint social groups. Finally, GED method is run with different user importance measures.

| opiekun pracy dyplomowej | ................................................. | ....................... | ....................... |
|---|---|---|---|
| | *Tytuł/stopień naukowy/imię i nazwisko* | *ocena* | *podpis* |

*Do celów archiwalnych pracę dyplomową zakwalifikowano do:**
*a) kategorii A (akta wieczyste)*
*b) kategorii BE 50 (po 50 latach podlegające ekspertyzie)*
** niepotrzebne skreślić*

Wrocław 2011

# TABLE OF CONTENT

# Abstract

The continuous interest in the social network area contributes to the fast development of this field. New possibilities of obtaining and storing data allows for more and more deeper analysis of the network in general, as well as groups and individuals within it. Especially interesting is studying the dynamics of changes in social groups over time. Having such knowledge ones may attempt to predict the future of the group, and then manage it properly in order to achieve presumed goals. Such ability would be a powerful tool in the hands of human resource managers, personnel recruitment, marketing, etc.

The thesis presents a new method for exploring the evolution of social groups, called Group Evolution Discovery (GED). Next, the results of its use are provided together with comparison to two other algorithms in terms of accuracy, execution time, flexibility and ease of implementation. Moreover, the method was evaluated with various measures of user importance within a group.

Obtained results suggest that GED is the best method for analyzing social group dynamics.

# Streszczenie

Nieustające zainteresowanie tematem sieci społecznych przyczynia się do szybkiego rozwoju tej dziedziny nauki. Nowe sposoby pozyskiwania i magazynowania danych pozwalają na coraz to głębszą analizę sieci jako ogółu, a także grup oraz jednostek w niej występujących. Szczególnie interesujące jest badanie dynamiki zmian zachodzących w grupach społecznych na przestrzeni czasu. Mając taką wiedzę można próbować przewidzieć przyszłość grupy, a następnie odpowiednio nią kierować aby osiągnąć założone cele. Taka umiejętność byłaby potężnym narzędziem w rękach osób zajmujących się zarządzaniem zasobami ludzkimi, doborem personelu, marketingiem, itp.

W pracy przedstawiono nową metodę do odkrywania ewolucji grup społecznych nazwaną Group Evolution Discovery (GED). Następnie pokazano wyniki jej użycia oraz porównano z dwoma innymi algorytmami pod kątem dokładności i szybkości działania, a także elastyczności i łatwości implementacji. Ponadto, metoda została sprawdzona z różnymi miarami ważności użytkowników w grupie społecznej.

Otrzymane wyniki sugerują, że GED jest najlepszą metodą do badania dynamiki grup społecznych.

# 1. Introduction

Social network in a simplest form is a social structure consisting of units that are connected by various kinds of relations like friendship, common interest, financial exchange, dislike, knowledge or prestige [27]. The easiest way to present social network in a mathematical way is graph representation where members are nodes of the graph and relations are edges between those nodes.

Social Network Analysis (SNA), which focuses on understanding the nature and consequences of relations between individuals or groups [61], [67] has become progressively attractive area within the social sciences for investigating human and social dynamics. The earliest basic text known of dealing exclusively with social network analysis is Knoke and Kuklinski’s Network Analysis, published in 1982 [50]. The development of SNA is so fast that the publications on methods and applications for analyzing social networks are updated almost every year. [29], [7].

Changes in technology and society creates a powerful mix of forces that will revolutionize the way all businesses – not just media companies – act, produce goods, and relate to customers [3]. There are plenty of reasons why SNA area should be examined, e.g. SNA can be used to help companies adapt to rapid economic changes [62], find key target markets, build up harmonious and successful project teams [10], help people find jobs [24], and more.

Social networks are dynamic by nature. A dynamic network consists of relations between members that evolves over time. Although, the idea is very simple and intuitive, tracking changes over time, especially changes within social groups is still uncharted territory on the social network analysis map. There are only a few methods dealing with this problem, and the need for more methods is tremendous.

This thesis presents new method for discovering group evolution in the social network. The method is evaluated on the email communication data in order to show its features and usefulness in the social network analysis area. The first results of the method were already presented in [70] and [71].

## 1.1. Aim and Objectives

The goal with this project is to identify and analyze the changes occurring in the social groups over the time. Additional objectives are:

1. to conduct research in literature on existing methods for detecting groups within the social network,
2. to extract social groups from large email communication dataset,
3. to conduct research in literature on existing methods for tracking group evolution,
4. to develop new algorithm for group evolution discovery working on both overlapping and disjoint group,
5. to identify and analyze changes occurring in social groups,
6. to prepare and conduct experiments which compare the new algorithm with existing ones.

## 1.2. Research Questions

The thesis addressed following research questions:

1. Which methods for detection groups within social network, methods based on fast modularity or methods based on cliques, are faster for the large datasets?
2. How changes of a social group over time can be noticed and evaluated?
3. What are the most common event types occurring in social group evolution?

4. What is the difference between various methods for tracking group evolution?
5. Which methods for discovering group evolution can be successfully used on overlapping groups?

## 1.3. Chapters Content

The rest of the thesis is organized as follows. Chapter 2 describes general concept of the social network and basis of the SN theory together with notation and representation of social group, also description of the social network analysis and measures used in the SNA and presentation of the temporal social network is provided. Chapter 3 describes the most common and valuable methods for group extraction and methods for quantifying group evolution in the social network. Chapter 4 presents the idea of new method for tracking group evolution preceded by theoretical basis, such as event types or inclusion measure required to understand the algorithm. Formula and pseudo-code is also provided in this section. Chapter 5 includes general scheme of all GED Platform's modules and detailed description of their tasks. Chapter 6 describes the email communication data used in the study together with data pre-processing needed to conduct the experiments. In Chapter 7 evaluation of the Group Evolution Discovery method focused on accuracy and flexibility is presented. Other aspects, such as execution time, ease of implementation and design are also mentioned. Moreover, exhaustive comparison with two other methods for tracking group evolution is provided. Chapter 8 includes outcomes from running experiments with answers for research questions. Additionally, development direction of GED method is revealed. Chapter 9 presents a lists of tables and figures occurring in the thesis, and finally Chapter 10 contains sorted list of the literature used in the thesis.

# 2. Social Network

## 2.1. General Concept of Social Network

There is no universally acceptable definition of the social network. Network analysed in this thesis can be described as set of actors (network nodes) connected by relationships (network edges). Many researchers proposed their own concept of social network [25], [61]. [67]. [68]. Social networks, as an interdisciplinary domain, might have different form: corporate partnership networks (law partnership) [40], scientist collaboration networks [48], movie actor networks, friendship network of students [4], company director networks [57], sexual contact networks [44], labour market [42], public health [8], psychology [51], etc.

The easiest to investigate, social networks, are online social networks [12], [20], web-based social networks [23], computer-supported social networks [69] or virtual social networks. The reason for this is simple and continuous way to obtain data from which those social networks can be extracted. Depending on the type of social network, data can be found in various places, e.g.: bibliographic data [21], blogs [2], photos sharing systems like Flickr [32], e-mail systems [65], [30], telecommunication data [6], [33], social services like Twitter [28] or Facebook [17], [64], video sharing systems like YouTube [11], Wikipedia [67] and more. Obtaining data from mentioned "data sources" allows to explore more than single social network in specific snapshot of time. Using proper techniques it is possible to evaluate changes occurring in social network over time. Especially interesting is following changes of social groups (communities) extracted from social networks.

## 2.2. Notation and Representation of Social Group

As there is no definition of the social network, there is also no common definition of the groups (communities) in social networks [18], [54]. Several different definitions are used, sometimes they are even simplified just to criteria for existence of the group [14], [19], [35]. Biologists described group as a cooperating entities, existing in the same environment. For sociologists community is a group of units sharing common area. Both definitions are focused on location of a members of a group. However, caused by fast propagation of the Internet, community is no more associated with geographical position. A general concept of a social group assumes that community is a set of units in given population (social network), who collaborate together more often than with other units of this population (social network). This general idea can be easily moved to the graph theory, where social network is represented as a graph and a community is a set of nodes (vertices) with high density of links (edges) within community, and lower density of a links directed outside the community. Moreover, communities can also be algorithmically determined, as the output of the specific clustering algorithm [43]. In this thesis, such a definition will be used, i.e. a group $G$ extracted from the social network $SN(V,E)$ is a subset of vertices from $V$ ($G \subseteq V$), extracted using any community extraction method (clustering algorithm).

## 2.3. Social Network Analysis

The term social networks have been used for the first time in the middle of 1950, but only in the 1980s researchers began to explore social relationships. Since then social network analysis (SNA) becomes necessary in an increasing number of application domains, including community discovery (as formation and evolution), social dynamics (as consensus, agreement and uniformity), recommendation systems and so on [22].

General idea of social network analysis is projecting and measuring of relationships and flows between people, communities, institutions, computers, web sites, and other knowledge processing units. SNA provides both a visual and a mathematical analysis of units relationships [36].

While performing social network analysis four main tasks can be observed. First step is selection of a sample which will be analyzed. Then the data can be collected using any existing method for collecting data, e.g. interviews, questionnaires, observation. There are two types of data that might be investigated, members and relations between them. Third step in SNA is choice and implementation of a social network analysis method. There are three approaches to the analyzing procedure (Figure 2.1):

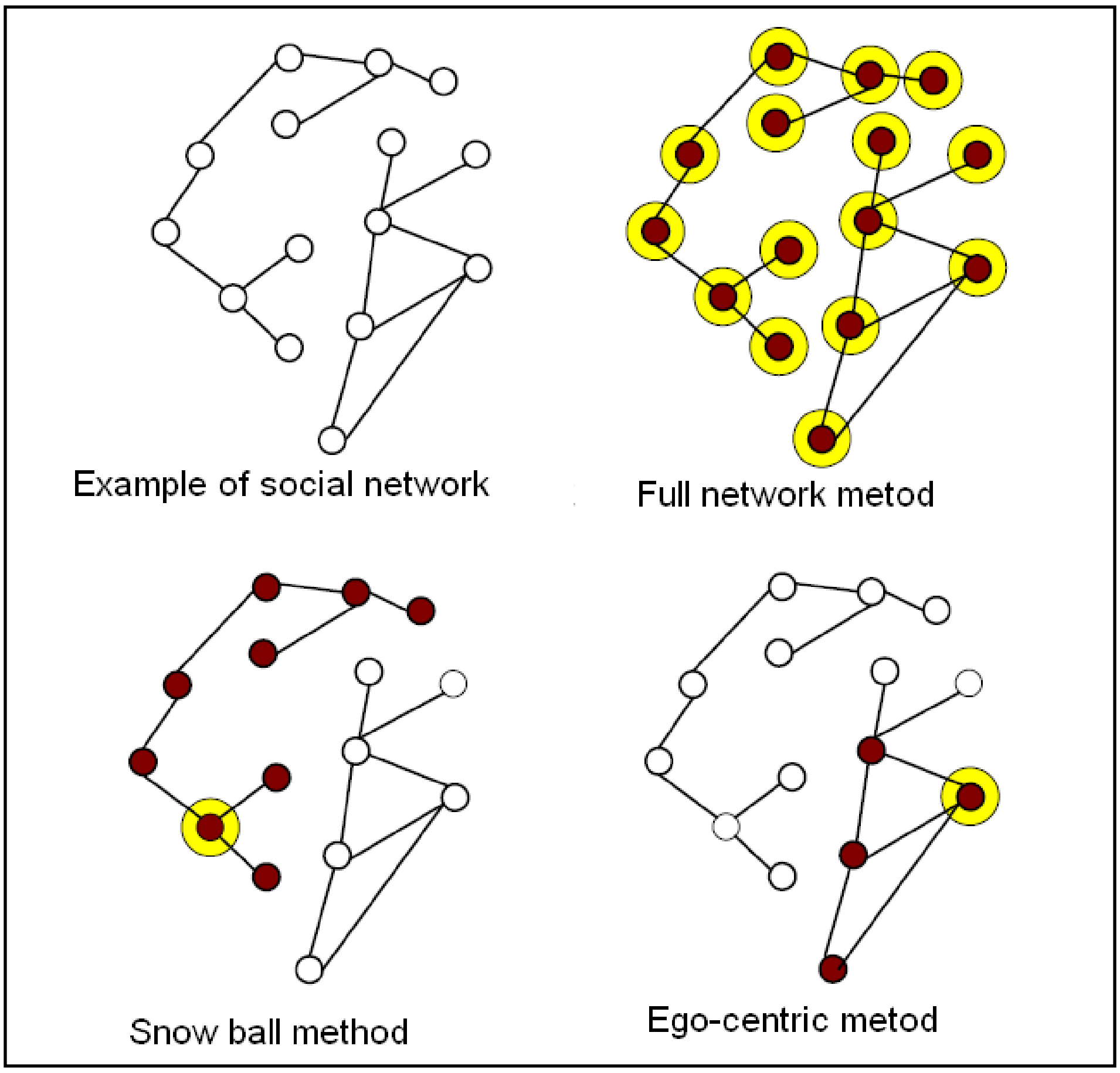

Figure 2.1. Visualisation of social network analysis methods (figure from [31]).

- **Full network methods** are collecting and analyzing data about the whole network. None of the members nor relationships is omitted. This approach is the most accurate but also the most expensive when it comes to computational cost or time needed for processing. Another inconvenience may be problem with collecting data for entire network [25].
- **Snowball methods** starts with single local member or small set of members and follow its relations in order to reach another members. Ones the method reach them the whole process is repeated until all members are investigated or the predefined number of iteration is exceeded. This method works very well for finding well connected groups in big networks but it also has some disadvantages. The method can omit members who are isolated or loosely connected, or in worst case the method can end on the first member because of lack in relations [25].

- **Ego-centric methods** are focused only on the single member and its neighbourhood (and also on relations between them). This method is useful for analyzing the local network and what influence has this network on considered member. Because of local character the method is very fast and computationally efficient [25].

The final step in social network analysis is drawing conclusions [20].

There are plenty of reasons why SNA area should be investigated, e.g. SNA can be used to identify target markets, create successful project teams and serendipitously identify unvoiced conclusion [10].

## 2.4. Measures in Social Network Analysis

While analyzing social network it is sometimes crucial to investigate which member is the most powerful (central). Or, looking from another angle, how important is the specific member within the social group, which he belongs to. To do so, one of the measures listed below may be used.

### *2.4.1. Social Position*

Social position is a measure which express the user importance within social network and is calculated in the iterative way. The social position for network *SN*(*V*,*E*) is calculated as follows [45]:

$$SP_{n+1}(x) = (1-\varepsilon) + \varepsilon \cdot \sum_{y \in V} SP_n(y) \cdot C(y \to x) \tag{2.1}$$

where:

$SP_{n+1}(x)$ – the social position of member $x$ after the $n$+1$^{st}$ iteration,

$\varepsilon$ – the coefficient from the range (0;1),

$C(y \to x)$ – the commitment function which expresses the strength of the relation from $y$ to $x$,

$SP_0(x) = 1$ for each $x \in V$.

Characteristic for the social position is that it takes into account both, the value of social positions of member's $x$ relations and their commitment in connections to $x$. In general, the greater social position one have the more profitable this user is for the entire network [45]. Algorithm of a method can be easily presented as follows:

1. For each member in a network assign $SP_0 = 1$.
2. For each edge $e(x,y)$ in a network recalculate SP of $y$ according to: $SP_n(y) = SP_n(y) + SP(x)_{n-1} \cdot C(x \to y)$
3. For each member in a network recalculate *SP* according to: $SP_n(x) = (1-\varepsilon) + \varepsilon \cdot SP_n(x)$
4. Repeat steps 2 and 3 until gain in *SP* for each member in a network is below presumed threshold.

Directed social network presented in Figure 2.2 contains commitment values between members. Based on these values and coefficient value $\varepsilon = 0{,}5$ social position of members is calculated in Table SN1. Each column represents one iteration of algorithm. Calculations stops when difference between successive iterations is 0,01 or lower. The final value of social position determines the rank of particular member within examined social network, the higher social position the higher position in the ranking. In the example illustrated in Figure 2.2 member C has the highest *SP* (and rank) due to the number of relations and their high commitment value from other members. Member D, in turn, has second place in the ranking as a result of just one relation, but it is relation from the most important member in the network. As easily seen in Table 2.1 the algorithm found the final ranking in the second iteration. In general, the smaller network the less iterations needed to calculate social positions.

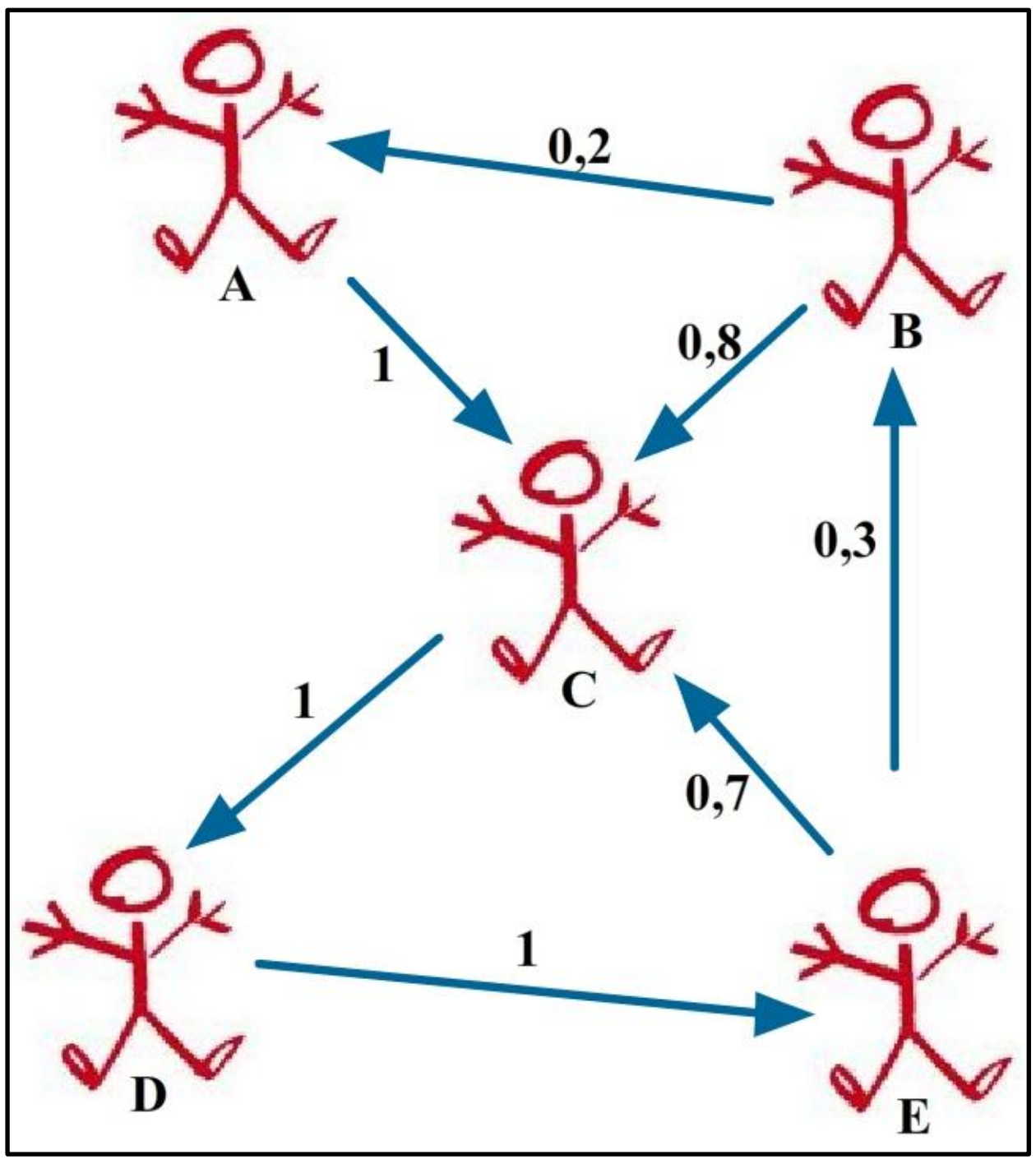

Figure 2.2. Members of a directed social network with assigned commitment values.

| **Member** | $SP_0$ | $SP_1$ | $SP_2$ | $SP_3$ | $SP_4$ | $SP_5$ | $SP_6$ | $SP_7$ | **Rank** |
|---|---|---|---|---|---|---|---|---|---|
| **A** | 1 | 0,60 | 0,565 | 0,565 | 0,565 | 0,568 | 0,567 | 0,566 | **5** |
| **B** | 1 | 0,65 | 0,650 | 0,650 | 0,678 | 0,665 | 0,665 | 0,667 | **4** |
| **C** | 1 | 1,75 | 1,410 | 1,393 | 1,458 | 1,440 | 1,434 | 1,440 | **1** |
| **D** | 1 | 1 | 1,375 | 1,205 | 1,196 | 1,229 | 1,220 | 1,217 | **2** |
| **E** | 1 | 1 | 1 | 1,188 | 1,103 | 1,098 | 1,115 | 1,110 | **3** |

Table 2.1. Social position of a members in successive iterations of algorithm. Last column contains ranking of a members in social network.

### *2.4.2. Centrality Degree*

The way of calculating centrality degree is very simple and intuitive. It is the number of direct connections of member *x* with other members [45]:

$$CD(x) = \frac{d(x)}{m-1} \tag{2.2}$$

where:
$d(x)$ – the number of members which are directly connected to member *x*,
*m* – the number of members in a network.

### *2.4.3. Centrality Closeness*

In a centrality closeness the member's location within the network is more important than the number of connections with other members (like in a centrality degree measure). A centrality closeness measure determines how close the member is to all other members in a network and counts how quick this member can get in touch with other members[45]:

$$CC(x) = \frac{m-1}{\sum_{\substack{y \neq x \\ y \in V}} c(x,y)} \tag{2.3}$$

where:

$c(x,y)$ – a function describing the distance between members *x* and *y*,

*m* – the number of members in a network.

### *2.4.4. Centrality Betweenness*

A centrality betweenness focuses on how many times member is located between two other members and how often the path goes through this member. Importance of the members with high centrality betweenness lies on fact that other members are connected with each other only by them. The measure is determined using [45]:

$$CB(x) = \frac{\sum_{\substack{x \neq y \neq z \\ x,y \in V}} \frac{b_{xy}(z)}{b_{xy}}}{m-1} \tag{2.4}$$

where:

$b_{xy}(z)$ – the number of shortest paths from *x* to *y* that goes through *z*,

$b_{xy}$ – the number of all paths from *x* to *y*,

*m* – the number of members in a network.

## 2.5. Temporal Social Network

Temporal social network *TSN* is a list of succeeding timeframes (time windows) *T*. Each timeframe is in fact one social network *SN*(*V*,*E*) where *V* is a set of vertices and *E* is a set of directed edges $< x, y >: x, y \in V,\ x \neq y$

$$\begin{aligned} &TSN = < T_1, T_2, ...., T_m >, \quad m \in N \\ &T_i = SN_i(V_i, E_i), \quad i = 1,2,...,m \\ &E_i = < x, y >: x, y \in V_i, x \neq y \quad i = 1,2,...,m \end{aligned} \tag{2.5}$$

Example of a temporal social network is presented in Figure 2.3. TSN consists of five timeframes, and each timeframe is social network created from data gathered in particular interval of time. In the simplest case one interval starts when previous interval ends, but based on author’s needs intervals may overlap by a set of time or even contain full history of previous timeframes.

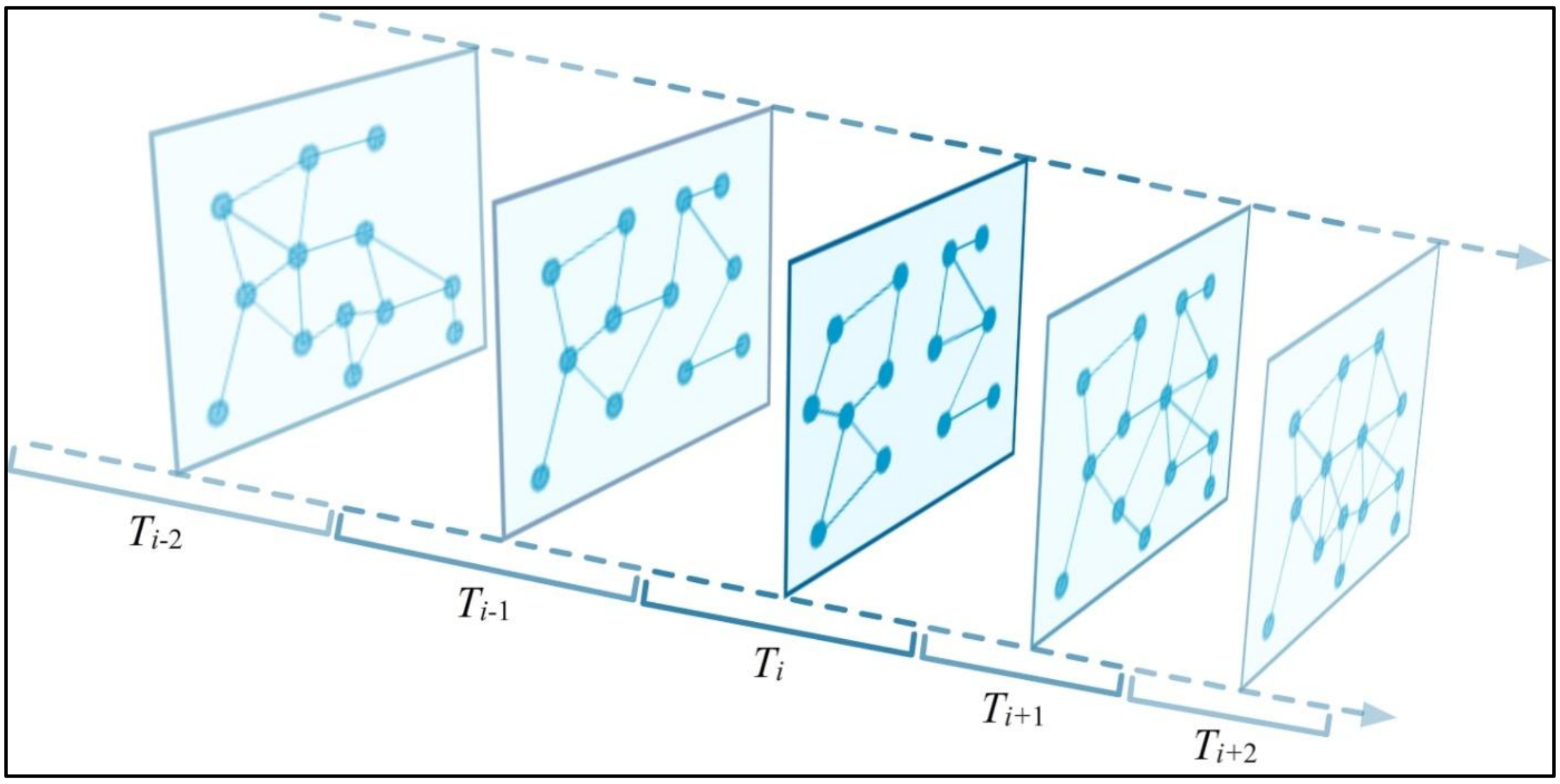

Figure 2.3. Example of temporal social network consisting of five timeframes.

# 3. Related Work

## 3.1. Methods for Group Extraction

Methods for group extraction, also called community detection methods or grouping methods, are the first step in analyzing social networks. The aim of these methods is to identify (extract) groups within a social network by only using the information contained in the network's graph. Two main types of the community detection methods can be distinguished, the one which assign each member to a single group, and the one which allows members to be the part of more than one community. The groups extracted with the first type of methods are called disjoint groups (they do not share any nodes), while the communities obtained by utilizing the second type of methods are called overlapping groups (they do share nodes). In further sections the most common and valuable methods for group extraction in the social network are presented.

### *3.1.1. Clique Percolation Method*

The clique percolation method (CPM) proposed by Palla et al. [54], [15] is the most widely used algorithm for extracting overlapping communities. The CPM method works locally and its basic idea assumes that the internal edges of a group has a tendency to form cliques as a result of high density between them. Oppositely, the edges connecting different communities are unlikely to form cliques. A complete graph with $k$ members is called $k$-clique. Two $k$-cliques are treated as adjoining if number of shared members is $k$–1. Lastly, a $k$-clique community is the graph achieved by the union of all adjoining $k$-cliques [1]. Such a assumption is made to represent fact that it is crucial feature of a group that its nodes can be attained through densely joint subsets of nodes. Algorithm works as follows:

1. All cliques are found for different values of $k$.
2. A square matrix $M = n \times n$, where n is the number of cliques found, is created. Each cell $[i, j]$ contains number of nodes shared by cliques $i$ and $j$.
3. All cliques of size equal or greater than $k$ are selected and between cliques of the same size connections are found in order to create a $k$-clique chain.

Palla et al. proposing their method aim for algorithm which is not too rigorous, takes into account the density of edges, works locally, and allows nodes to be a part of several groups. All the requirements were fulfilled, moreover Palla and co-workers [18] implemented CPM algorithm in software package called CFinder, which is freely available at [26].

### *3.1.2. Fast Modularity Optimization*

The method by Blondel et al. [6] is designed to deal with the large social networks. It provides good quality of the extracted disjoint groups in low computation time, what is more a complete hierarchical community structure is also supplied. In the first step algorithm creates a different community for each member of the network. Then, repeating iteratively members are moved to neighbours' communities, but only if such action will improve the modularity of the considered group. Gain in modularity $\Delta Q$ obtained by adding node $i$ into a community $C$ is calculated as follows:

$$\Delta Q = \left[ \frac{\sum_{in} + k_{i,in}}{2m} - \left( \frac{\sum_{tot} + k_i}{2m} \right)^2 \right] - \left[ \frac{\sum_{in}}{2m} - \left( \frac{\sum_{tot}}{2m} \right)^2 - \left( \frac{k_i}{2m} \right)^2 \right] \tag{3.1}$$

where:
$\sum_{in}$ – the sum of the weights of the links inside community *C*,
$\sum_{tot}$ – the sum of the weights of the links incident to nodes in community *C*,
$k_i$ – the sum of the weights of the links incident to node *i*,
$k_{i,in}$ – the sum of the weights of the links from node *i* to nodes in community *C*,
*m* – the sum of the weights of all the links in the network.

Algorithm stops when none of the members cannot increase the modularity of its neighbours' group. Algorithm, step by step, is presented below.

1. Each node is assigned to separate group.
2. Each node is removed from its group and added to the neighbour's group, gain in modularity is counted and node stays in group where gain is the biggest. If the gain in modularity is below zero for all neighbours' groups, the node goes back to its original group.
3. Step 2 is repeated until modularity cannot be improved any more.
4. New network is created, where groups are represented by super-nodes. Super-nodes are connected if there is at least one connection between groups represented by super-nodes. The weight of the connection equals sum of weights of connections between groups.
5. Steps 1 – 4 are repeated until the network consist of one super-node.

The biggest advantages of the method are intuitive concept of grouping nodes, ease of implementation, extremely low computational cost, and unfolding hierarchical community structure. Method by Blondel et al. is implemented for example in a Workbench for Network Scientists (NWB) [49].

### *3.1.3. Algorithm of Girvan and Newman*

Method by Girvan and Newman [21] [46] is one of the best known algorithms for extracting disjoint groups. This method focuses on the edges which are least central in order to remove them from the network. To determine the weakest edges, those which are most "between" groups, Girvan and Newman used slightly modified betweenness centrality measure (mentioned in section 2.4.4.). The edges are iteratively removed from the network, based on the value of their betweenness. After each iteration betweenness of the edges affected by the removal is recalculated. Algorithm stops when there are no edges to remove, which means that all groups have been disjointed.

### *3.1.4. Radicchi et al. Method*

Radicchi et al. in [55] proposed faster version of Girvan-Newman method. A divisive algorithm requires the consideration of only local quantities. The authors used *edge-clustering coefficient* to single out edges connecting members belonging to different groups. Having the same accuracy as algorithm of Girvan and Newman, method by Radicchi et al. works much faster, allowing to investigate far bigger networks.

#### *3.1.5. Lancichinetti et al. Method*

Algorithm by Lancichinetti et al. [39] identifies the natural communities of the members based on their *fitness*. The fitness is calculated from the internal and external degrees of the members in communities. Counting fitness for every node in graph will cover it by the overlapping groups. Due to the parameter controlling the size of the communities there is a possibility to find hierarchical dependencies between groups. The method is very flexible, fitness function can be designed for particular type of network, e.g. weighted networks.

#### *3.1.6. iLCD Algorithm*

In order to detect communities, Cazabet et al. in [56] focused not only on edges and nodes within group, but also on its particular pattern of development. When new member appears in the network algorithm looks for groups which will suites new node. Suits in this case means that (1) the number of neighbours inside the community which new member can access with a path of length two or less is higher than the mean number of second neighbours within community, and (2) the number of neighbours inside the community which new member can access with a path of length two or less, by at least two different paths is greater than the mean number of robust second neighbours within community. The intrinsic Longitudinal Community Detection (iLCD) algorithm allows groups to overlap, which makes it optional for CPM method.

#### *3.1.7. Other Methods*

Apart from the most common, presented above, methods for detecting groups in a network, researchers developed many other, e.g. Fast greedy modularity optimization by Clauset, Newman and Moore [13], Markov Cluster Algorithm [66], Structural algorithm by Rosvall and Bergstrom [59], Dynamic algorithm by Rosvall and Bergstrom [60], Spectral algorithm by Donetti and Muñoz [16], Expectation-maximization algorithm by Newman and Leicht [47], Potts model approach by Ronhovde and Nussinov [58]. Most of them are analyzed and evaluated by Lancichinetti and Fortunato in [38].

## 3.2. Methods for Tracking Group Evolution

One aspect of the social network analysis is to investigate dynamics of a community, i.e., how particular group changes over time. To deal with this problem several methods for tracking group evolution have been proposed. Almost all of them as a input data needs the social network with communities extracted by one of the grouping methods. In a consequence specific methods for tracking evolution works better on disjoint groups or on overlapping groups. Further paragraphs provides the basic ideas behind the most popular methods for analyzing social group dynamics.

#### *3.2.1. Asur et al. Method*

The method by Asur et al. [5] has simple and intuitive approach for investigating community evolution over time. The group size and overlap are compared for every possible pair of groups in the consecutive timeframes and events involving those groups are assigned. When none of the nodes of community from timeframe *i* occur in following timeframe *i*+1, Asur et al. described this situation as dissolve of the group.

$$Dissolve\left(C_i^k\right) = 1 \text{ iff } \exists \text{ no } C_{i+1}^j \text{ such that } V_i^k \cap V_{i+1}^j > 1 \tag{3.2}$$

where:

$C_i^k$ – community number *k* in timeframe number *i*,

$V_i^k$ – the set of the vertex (nodes) of community number *k* in timeframe number *i*.

In opposite to dissolve, if none of the nodes of community from timeframe *i*+1 was present in previous timeframe *i*, group is marked as new born.

$$Form\left(C_{i+1}^k\right) = 1 \text{ iff } \exists \text{ no } C_i^j \text{ such that } V_{i+1}^k \cap V_i^j > 1 \tag{3.3}$$

Community continue its existence when identical occurrence of the group in consecutive timeframe is found.

$$Continue\left(C_i^k, C_{i+1}^j\right) = 1 \text{ iff } V_i^k = V_{i+1}^j \tag{3.4}$$

Situation when two communities from timeframe *i* joint together overlap with more than $\kappa\%$ of the single group in timeframe *i*+1, is called merge.

$$\begin{gathered} Merge\left(C_i^k, C_i^l, \kappa\right) = 1 \text{ iff } \exists\, C_{i+1}^j \text{ such that } \frac{\left|\left(V_i^k \cup V_i^l\right) \cap V_{i+1}^j\right|}{Max\left(\left|V_i^k \cup V_i^l\right|, \left|V_{i+1}^j\right|\right)} > \kappa\% \\ \text{and } \left|V_i^k \cap V_{i+1}^j\right| > \frac{\left|C_i^k\right|}{2} \text{ and } \left|V_i^l \cap V_{i+1}^j\right| > \frac{\left|C_i^l\right|}{2} \end{gathered} \tag{3.5}$$

Opposite case, when two groups from timeframe *i*+1 joint together overlap with more than $\kappa\%$ of the single group in timeframe *i*, is marked as split.

$$\begin{gathered} Split\left(C_i^j, \kappa\right) = 1 \text{ iff } \exists\, C_{i+1}^k, C_{i+1}^l \text{ such that } \frac{\left|\left(V_{i+1}^k \cup V_{i+1}^l\right) \cap V_i^j\right|}{Max\left(\left|V_{i+1}^k \cup V_{i+1}^l\right|, \left|V_i^j\right|\right)} > \kappa\% \\ \text{and } \left|V_{i+1}^k \cap V_i^j\right| > \frac{\left|C_{i+1}^k\right|}{2} \text{ and } \left|V_{i+1}^l \cap V_i^j\right| > \frac{\left|C_{i+1}^l\right|}{2} \end{gathered} \tag{3.6}$$

Authors of the method suggested 30% or 50% as a value for $\kappa$ threshold. Example of the events described by Asur et al. are presented in Figure 3.1. Communities $C_1^1$ and $C_1^2$ continue between timeframes 1 and 2, then merge into one community $C_3^1$ in timeframe 3. In timeframe 4 community $C_3^1$ splits into three groups $C_4^1$, $C_4^2$ and $C_4^3$, next in timeframe 5 new community $C_5^4$ forms and finally in timeframe 6 the biggest community $C_5^1$ dissolves.

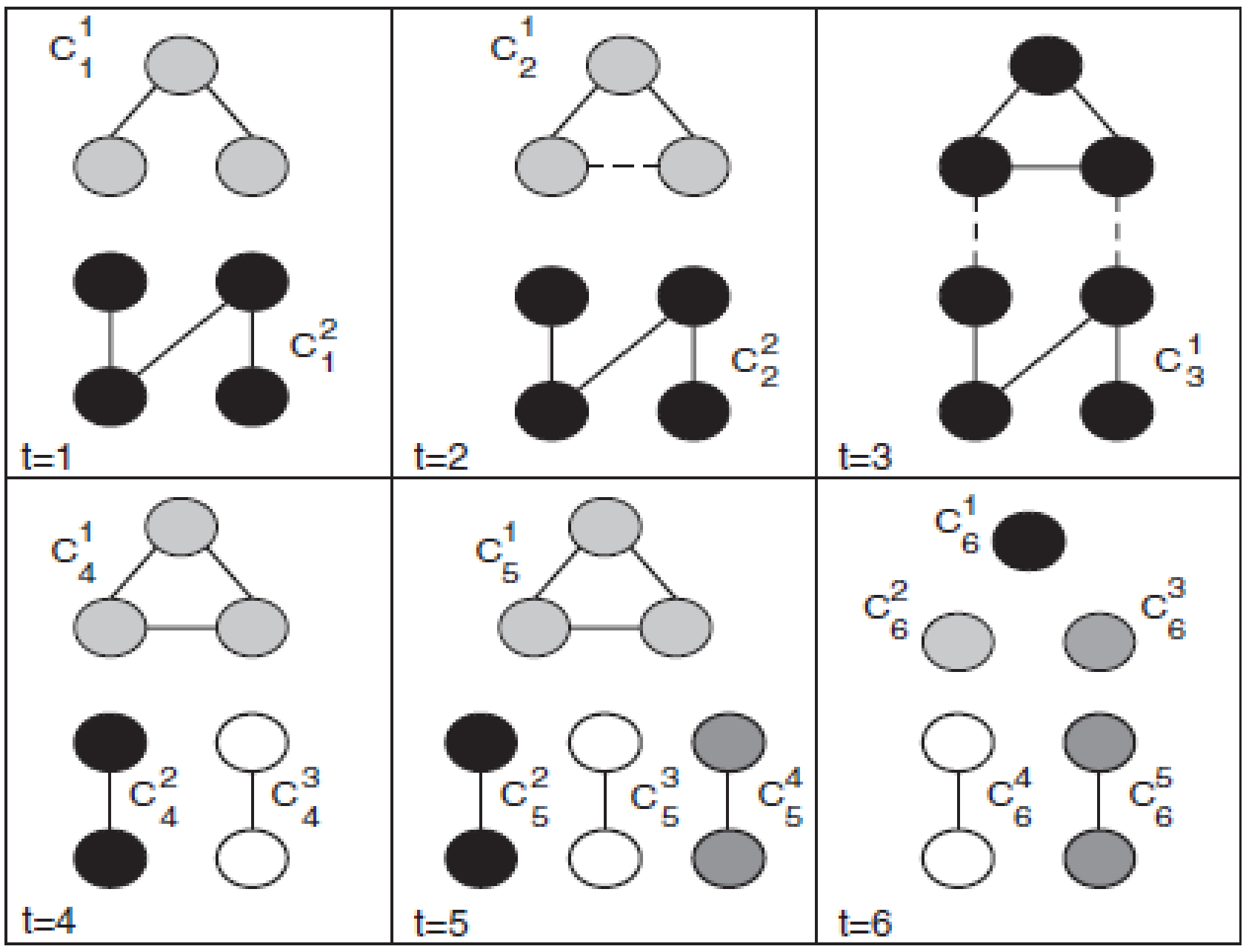

Figure 3.1. Possible group evolution by Asur et al. (figure from [5]).

Method provided by Asur et al. allows also to investigate behaviour of individual members in a community life. Node can appear in a network, disappear from a network, and also join and leave community.

Unfortunately, Asur et al. did not specify which method should be used for community detection, nor if method works for overlapping groups.

### *3.2.2. Palla et al. Method*

Palla et al. in their method [52], [53] used all advantages of the clique percolation method (described in section 3.1.1.) for tracking social group evolution. Social networks at two consecutive timeframes $i$ and $i$+1 are merged into single graph $Q(i, i+1)$ and groups are extracted using CPM method. Next, the communities from timeframes $i$ and $i$+1, which are the part of the same group from joint graph $Q(i, i+1)$, are considered to be matching i.e. community from timeframe $i$+1 is considered to be an evolution of community from timeframe $i$. It is common that more than two communities are contained in the same group from joint graph (Figure 3.2b and Figure 3.2c). In such a case matching is performed based on the value of their relative overlap sorted in descending order. The overlap is calculated as follows:

$$O(A,B) = \frac{|A \cap B|}{|A \cup B|} \tag{3.7}$$

where:

$|A \cap B|$ – the number of common nodes in the communities $A$ and $B$,

$|A \cup B|$ – the number of nodes in the union of the communities $A$ and $B$.

However, the authors of the method did not explain how to chose the best match for the community, which in next timeframe has the highest overlap with two different groups.

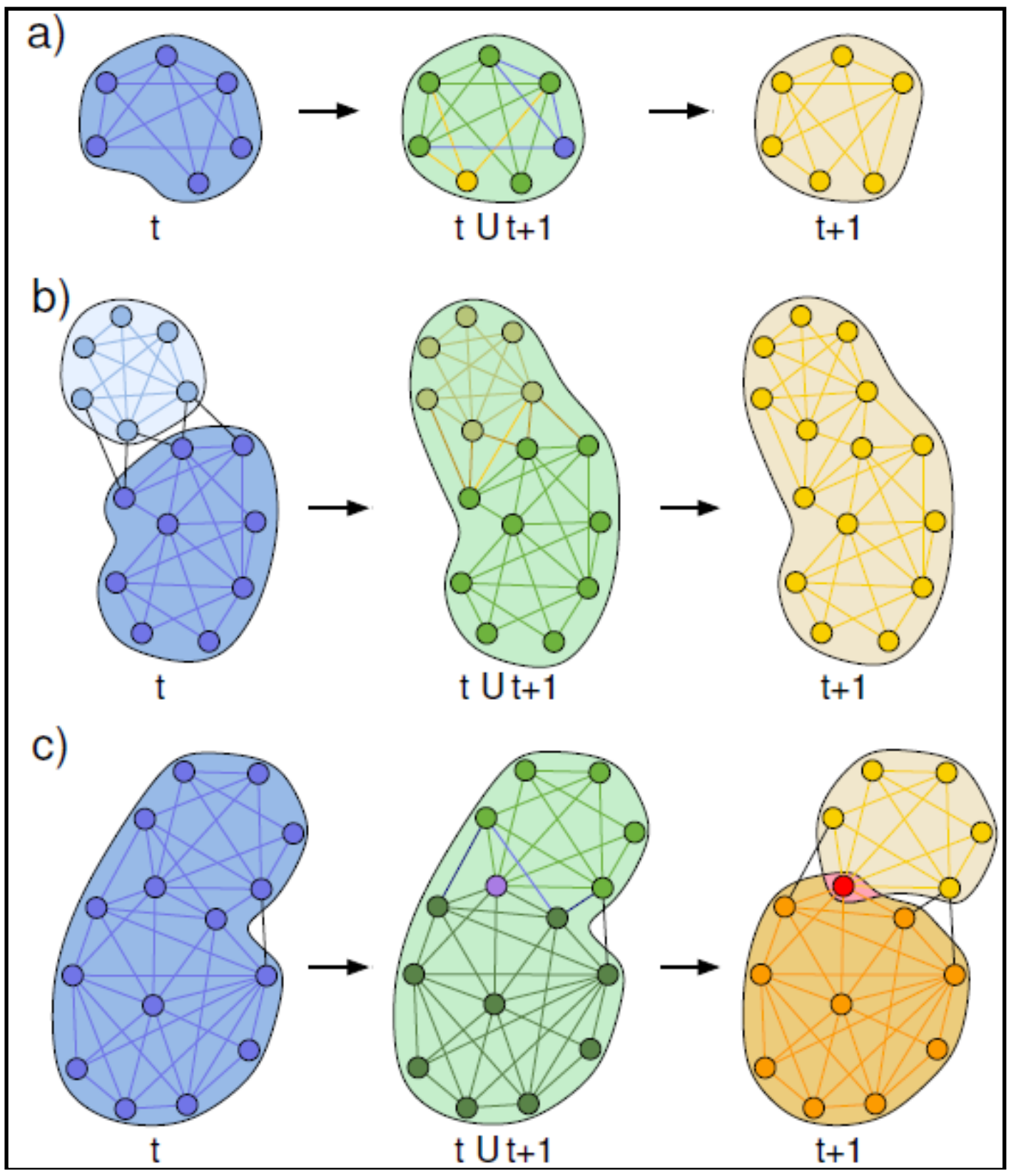

Figure 3.2. Most common scenarios in the group evolution by Palla et al.. The groups at timeframe *t* are marked with blue, the groups at timeframe *t*+1 are marked with yellow, and the groups in the joint graph are marked with green. a) a group continue its existence, b) the dark blue group swallows the light blue, c) the yellow group is detached from the orange one (figure from [53]).

Palla et al. proposed several event types between groups: growth, contraction, merge, split, birth and death, but no algorithm to assign them is provided. The biggest disadvantage of the method by Palla et al. is that it has to be run with CPM, no other method for community evolution can be used. Despite some lacks, the method is considered the best algorithm tracking evolution for overlapping groups.

### *3.2.3. Chakrabarti et al. Method*

Chakrabarti et al. in their method [9] presented original concept for the identifying group changes over time. Instead of extracting communities for each timeframe and matching them, the authors of the method introduced the *snapshot quality* to measure the accuracy of the partition $C_t$ in relation to the graph formation at time *t*. Then the *history cost* measures difference between partition $C_t$ and partition at the previous timeframe $C_{t\text{-}1}$. The total worth of $C_t$ is the sum of snapshot quality and history cost at each timeframe. Most valuable partition is the one with high snapshot quality and low history cost. To obtain $C_t$ from $C_{t\text{-}1}$, Chakrabarti et al. use relative weight *cp* (tuned by user) to minimize difference between snapshot quality and history cost. Chakrabarti et al. did not mention if method work for overlapping groups.

### *3.2.4. Kim and Han Method*

Kim and Han in their method [34] used *links* to connect nodes at timeframe *t*–1 with nodes at timeframe *t*, creating *nano-communities*. The nodes are connected to their future occurrences and to their future neighbours. Next, the authors analyzed the number and density of the links to judge which case of relationship occurs for given nano-community. Kim and Han stated most common changes, which are: evolving, forming and dissolving. Evolving of a group can be distinguished into three different cases: growing, shrinking and drifting. Community $C_t$ grows between timeframes $t$ and $t$+1 if there is a group $C_{t+1}$ in the following timeframe containing all nodes of $C_t$. Group $C_{t+1}$ may, of course, contain additional nodes, which are not present in $C_t$. In opposite, community $C_t$ shrinks between timeframes $t$ and $t$+1 when there is a group $C_{t+1}$ in the next timeframe which all nodes are contained in $C_t$. Finally, group $C_t$ is drifting between timeframes $t$ and $t$+1 if there is group $C_{t+1}$ in the following timeframe which has at least one node common with $C_t$. Kim and Han did not specify if the method is designed for overlapping or disjoint groups, but the drifting event suggest that method will not work correctly for overlapping groups.

### *3.2.5. FacetNet*

Lin et al. used evolutionary clustering to create FacetNet [41], a framework allowing members to be a part of more than one community at given timeframe. In contrast to Chakrabarti et al. method, Lin et al. used the *snapshot cost* and not the snapshot quality to calculate adequate of the partition to the data. Kullback-Leibler method [37] has been used for counting snapshot cost and history cost. Based on results of FacetNet it is easier to follow what happens with particular nodes, rather than what happens with a group in general. The algorithm is not assigning any events, but user can analyze results and assign events on his own. Unfortunately, FacetNet is unable to catch forming and dissolving events.

### *3.2.6. GraphScope*

Sun et al. presented parameter-free method called GraphScope [63]. At the first step partitioning is repeated until the smallest *encoding cost* for a given graph is found. Subsequent graphs are stored in the same segment $S_i$ if encoding cost is similar. When examined graph $G$ has higher encoding cost than encoding cost of segment $S_i$, graph $G$ is placed to segment $S_{i+1}$. Jumps between segments marks change-points in graph evolution over time. The main goal of this method is to work with a streaming dataset, i.e. method has to detect new communities in a network and decide when structure of the already existing communities should be changed in the database. Therefore, to adapt GraphScope for tracking group evolution, some extensions are needed.

# 4. Group Evolution Discovery Method

The small number of algorithms for tracking community evolution, as well as their low flexibility and accuracy suggest a gap in the knowledge. Therefore, in this thesis, the new method for the group evolution discovery, called GED, is proposed. Further sections presents the particular elements of the method and explains their usefulness.

## 4.1. Community Evolution

Evolution of particular social community can be represented as a sequence of events (changes) following each other in the successive timeframes within the temporal social network. Possible events in social group evolution are:

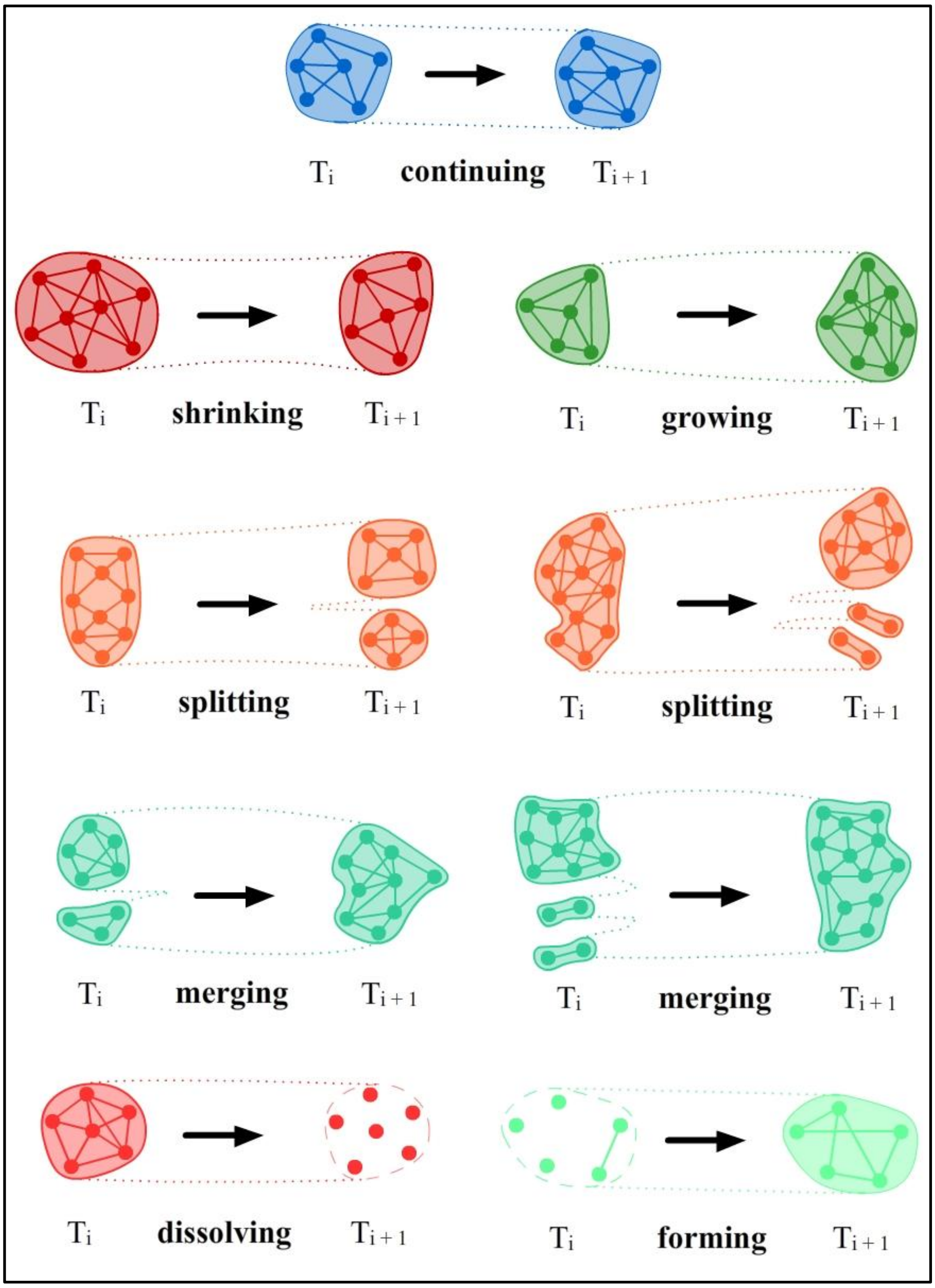

Figure 4.1. The events in community evolution.

- *Continuing* (stagnation) – the community continue its existence when two groups in the consecutive time windows are identical or when two groups differ only by few nodes but their size is the same. Intuitively, when two communities are so much similar that it is hard to see the difference.
- *Shrinking* – the community shrinks when some members has left the group, making its size smaller than in the previous time window. Group can shrink slightly, losing only few nodes, or greatly, losing most of its members.
- *Growing* (opposite to shrinking) – the community grows when some new members have joined the group, making its size bigger than in the previous time window. A group can grow slightly as well as significantly, doubling or even tripling its size.
- *Splitting*– the community splits into two or more communities in the next time window when few groups from timeframe $T_{i+1}$ consist of nodes of one group from timeframe $T_i$. Two types of splitting can be distinguished: (1) equal, which means the contribution of the groups in split group is almost the same and (2) unequal, when one of the groups has much greater contribution in the split group. In second case for the biggest group the splitting might looks similar to shrinking.
- *Merging* (reverse to splitting) – the community has been created by merging several other groups when one group from timeframe $T_{i+1}$ consist of two or more groups from the previous timeframe $T_i$. Merge, just like the split, might be (1) equal, which means the contribution of the groups in merged group is almost the same, or (2) unequal, when one of the groups has much greater contribution into the merged group. In second case for the biggest group the merging might looks similar to growing.
- *Dissolving* happens when a community ends its life and does not occur in the next time window, i.e., its members have vanished or stop communicating with each other and scattered among the rest of the groups.
- *Forming* (opposed to dissolving) of new community occurs when group which has not existed in the previous time window $T_i$ appears in next time window $T_{i+1}$. In some cases, a group can be inactive over several timeframes, such case is treated as dissolving of the first community and forming again of the, second, new one.

The examples of events described above are illustrated in Figure 4.1.

The easiest way to track whole evolution process for the particular community is to combine all changes during its lifetime to a single graph (Figure 4.2) or table (Table 4.1).

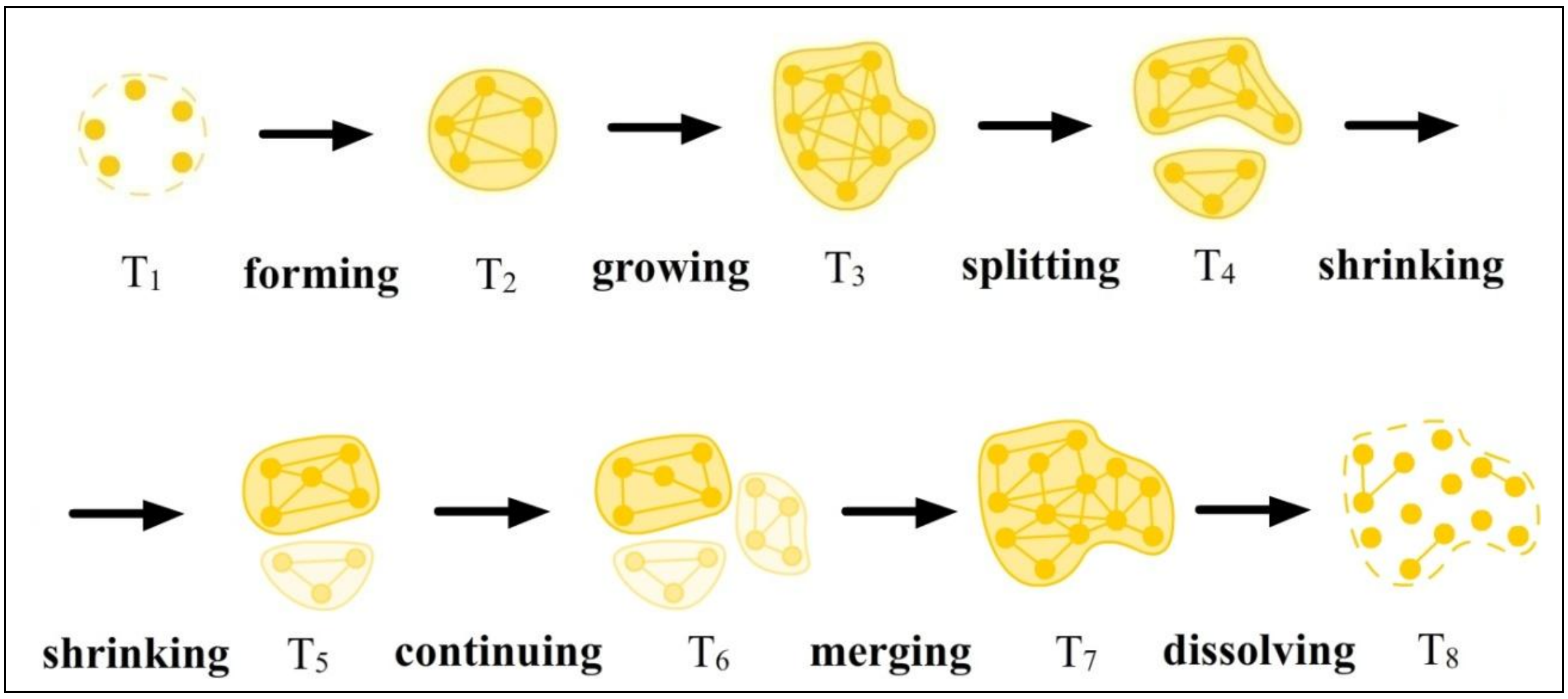

Figure 4.2. Evolution of the single community presented on a graph.

In the examples presented in Figure M2. and in Table M1. the network consists from eight timeframes. Group $G_1$ forms in $T_2$, which means that members of $G_1$ have no relations in $T_1$ or relations are rare. Next, by gaining four new nodes, community grows in $T_3$. In following timeframe $T_4$ group $G_1$ splits into $G_2$ and $G_3$. By losing one node, group $G_2$ shrinks in $T_5$ while group $G_3$ remains unchanged. Then new group $G_4$ forms in $T_6$, while both communities $G_2$ and $G_3$ continue their existence. In timeframe $T_7$ all groups merges into one community $G_5$ but in last timeframe $T_8$ group dissolves preserving only few relations between its members.

| **Event** | **$T_2$** | **Event** | **$T_3$** | **Event** | **$T_4$** | **Event** | **$T_5$** | **Event** | **$T_6$** | **Event** | **$T_7$** | **Event** |
|---|---|---|---|---|---|---|---|---|---|---|---|---|
| form | $G_1$ | growth | $G_1$ | split | $G_2$ | shrink | $G_2$ | continue | $G_2$ | merge | $G_5$ | dissolve |
| form | $G_1$ | growth | $G_1$ | split | $G_3$ | continue | $G_3$ | continue | $G_3$ | merge | $G_5$ | dissolve |
| - | - | - | - | - | - | - | - | form | $G_4$ | merge | $G_5$ | dissolve |

Table 4.1. Evolution of the communities presented in a table.

## 4.2. Inclusion Measure

To be able to track social community evolution, the groups from successive timeframes have to be matched into pairs. The most common and simplest approach is counting the overlapping of those groups:

$$O(G_1, G_2) = \frac{|G_1 \cap G_2|}{MAX(|G_1|, |G_2|)} \tag{4.1}$$

where:

$|G_1 \cap G_2|$ – the number of shared nodes.

$MAX(|G_1|, |G_2|)$ – the size of the bigger group.

However, overlap function can easily miss important relationships, e.g., when one group is small and another one is huge overlapping will be low and the methods for tracking evolution will ignore this pair of the groups. To avoid such a situations and to emphasize relations within the community a novel measure called *inclusion* is proposed. This measure allows to evaluate the inclusion of one group in another. Therefore, inclusion of group $G_1$ in group $G_2$ is calculated as follows:

$$I(G_1, G_2) = \frac{\overbrace{|G_1 \cap G_2|}^{\text{group quantity}}}{|G_1|} \cdot \underbrace{\frac{\sum_{x \in (G_1 \cap G_2)} SP_{G_1}(x)}{\sum_{x \in (G_1)} SP_{G_1}(x)}}_{\text{group quality}} \cdot 100\% \tag{4.2}$$

where:

$SP_{G_1}(x)$ – value of social position of the member $x$ in $G_1$.

The unique structure of this measure takes into account both the quantity and quality of the group members. The quantity is reflected by the first part of the inclusion measure ,i.e., what portion of $G_1$ members is shared by both groups $G_1$ and $G_2$, whereas the quality is expressed by the second part of the inclusion measure, namely what contribution of important members of $G_1$ is shared by both groups $G_1$ and $G_2$. It provides a balance between the groups which contain many of the less important members and groups with only few but key members.

The one might say that inclusion formula is “unfair” for not identical groups, because if community differ even by only one member, inclusion is reduced for not having all nodes and also for not having social position of those nodes. Indeed, it is slightly “unfair” (or rather strict), but using social position measure, which is calculated based on members’ relations, causes that inclusion focuses not only on nodes (members) but also on edges (relations) giving great advantage over overlapping measure.

Naturally, instead of social position (*SP*) any other measure which indicates user importance can be used e.g. centrality degree, closeness degree, betweenness degree, etc. But it is important that this measure is calculated for the group and not for social network in order to reflect node importance in community and not in the whole social network.

## 4.3. Algorithm

As mentioned before, the overlap measure has a tendency to missing important evolutions, therefore inclusion is counted for both groups separately. Then, even if the inclusion of huge group in small one is low, the opposite inclusion, the inclusion of small group in huge group can still have high value. In such a case, the method will not skip any meaningful evolutions.

Intuitively, between two groups $<G_1, G_2>$ only one event may occur, e.g. community $G_1$ cannot shrinks and merge into community $G_2$ at the same time. Of course one community in timeframe $T_i$ may have several events with different communities in $T_{i+1}$, e.g. $G_1$ can split into $G_2$ and $G_3$. Assigning events with GED method is based on the size of the communities and on the inclusion values of both groups, if at least one of the inclusions exceeds the *thresholds* set by the user, the event is assigned, (see Figure M3.). The exceptions are events forming and dissolving, which are assigned with special condition. In order to assign forming (dissolving) event members of a community cannot have relations in previous (next) timeframe or relations have to be rare, i.e. considered group must have very low inclusions level with all groups in previous (next) timeframe. In this thesis a very low level is regarded as a value below 10%, argumentation for that is presented in experimental section.

The user can set value of each threshold individually, $\alpha$ threshold is for inclusion of group $G_1$ in $G_2$, while $\beta$ threshold is for inclusion of $G_2$ in $G_1$. The value of thresholds has to be from range <0%, 100%>, however it is recommended to choose values above 50% to guarantee good inclusion of matching communities. An advantage of counting two inclusions instead of one was already provided, what is the profit of using two thresholds? Primarily, the method gains on flexibility and the user has possibility to obtain the results which he needs. The extensive explanation on setting value of thresholds and their influence on results are provided in experimental section of this thesis.

**_GED_ – Group Evolution Discovery Method**

**Input:**

*TSN* in which at each timeframe $T_i$ groups are extracted by any community detection algorithm; calculated any user importance measure.

1. For each pair of groups $<G_1, G_2>$ in consecutive timeframes $T_i$ and $T_{i+1}$ inclusion of $G_1$ in $G_2$ and $G_2$ in $G_1$ is counted according to equations (MW2).
2. Based on inclusion and size of two groups one type of event may be assigned:
   a. *Continuing*: $I(G_1,G_2) \geq \alpha$ and $I(G_2,G_1) \geq \beta$ and $|G_1| = |G_2|$

b. *Shrinking*: $I(G_1,G_2) \geq \alpha$ and $I(G_2,G_1) \geq \beta$ and $|G_1| > |G_2|$

OR

$I(G_1,G_2) < \alpha$ and $I(G_2,G_1) \geq \beta$ and $|G_1| \geq |G_2|$ and there is only one match (matching event) between $G_2$ and all groups in the previous timeframe $T_i$

c. *Growing*: $I(G_1,G_2) \geq \alpha$ and $I(G_2,G_1) \geq \beta$ and $|G_1| < |G_2|$

OR

$I(G_1,G_2) \geq \alpha$ and $I(G_2,G_1) < \beta$ and $|G_1| \leq |G_2|$ and there is only one match (matching event) between $G_1$ and all groups in the next timeframe $T_{i+1}$

d. *Splitting*: $I(G_1,G_2) \geq \alpha$ and $I(G_2,G_1) < \beta$ and $|G_1| \geq |G_2|$ and there is more than one match (matching events) between $G_2$ and all groups in the previous time window $T_i$

OR

$I(G_1,G_2) < \alpha$ and $I(G_2,G_1) \geq \beta$ and $|G_1| \geq |G_2|$ and there is more than one match (matching events) between $G_2$ and all groups in the previous time window $T_i$

e. *Merging*: $I(G_1,G_2) \geq \alpha$ and $I(G_2,G_1) < \beta$ and $|G_1| \leq |G_2|$ and there is more than one match (matching events) between $G_1$ and all groups in the next time window $T_{i+1}$

OR

$I(G_1,G_2) < \alpha$ and $I(G_2,G_1) \geq \beta$ and $|G_1| \leq |G_2|$ and there is more than one match (matching events) between $G_1$ and all groups in the next time window $T_{i+1}$

f. *Dissolving*: for $G_1$ in $T_i$ and each group $G_2$ in $T_{i+1}$ $I(G_1,G_2) < 10\%$ and $I(G_2,G_1) < 10\%$

g. *Forming*: for $G_2$ in $T_{i+1}$ and each group $G_1$ in $T_i$ $I(G_1,G_2) < 10\%$ and $I(G_2,G_1) < 10\%$

The scheme which facilitate understanding of the event selection for the pair of groups in the method is presented in Figure 4.3.

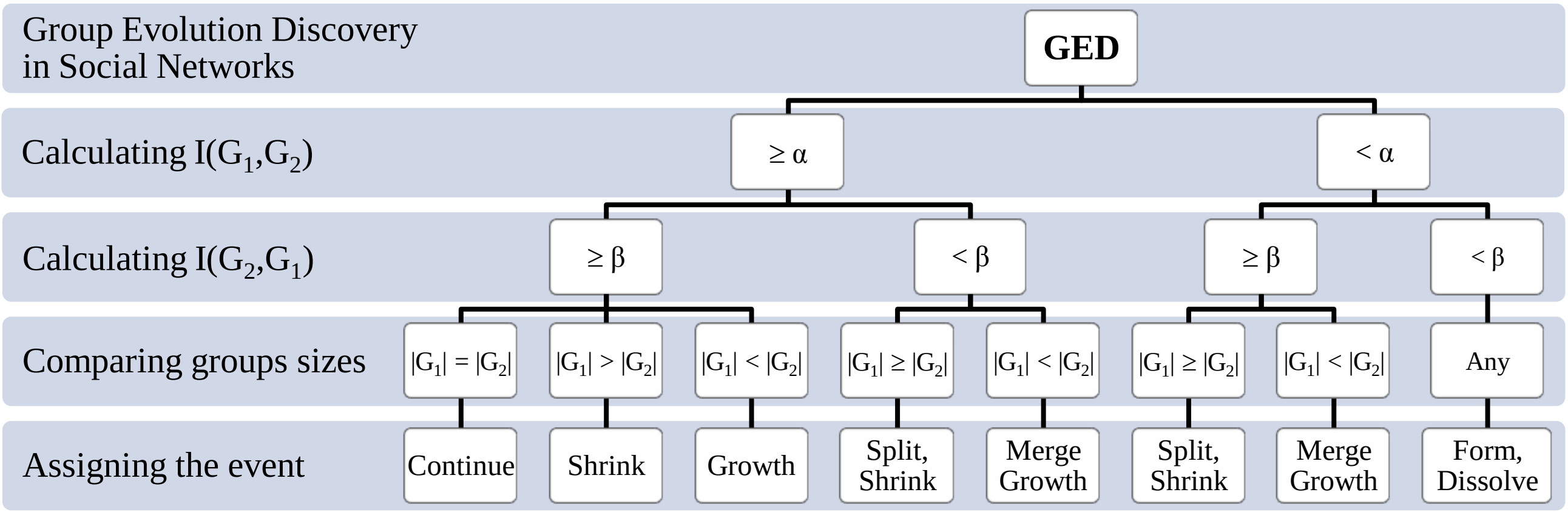

Figure 4.3. Decision tree for assigning the event type to the group.

Based on the list of extracted events, which have occurred for selected community between each two successive timeframes, the group evolution is created (Figure 4.2).

## 4.4. Pseudo-code

Pseudo-code of the algorithm can be implemented in any programming language, however the lowest execution time can be achieved with SQL languages, which are aimed for processing large datasets, e.g. T-SQL language.

***GED* – Group Evolution Discovery Method**

**Input:**

*TSN* in which at each timeframe $T_i$ groups are extracted by any community detection algorithm; calculated any user importance measure.

**Output:**

The list of communities matched into pairs with assigned event type and calculated inclusions.

```
begin
  for (each group in Ti) do
    begin
    for (each group in Ti+1) do
      begin
      calculate inclusions I(G1,G2) and I(G2,G1)
      assign the event based on Figure 4.3 and add matched pair to the list
      end;
    end;
    for (each pair on the list) update splitting/shrinking, merging/growing
      begin
      if (there is only one match between G2 and all groups in the previous (next)
        timeframe) set shrinking (growing)
      else set splitting (merging)
      end;
end.
```

# 5. Group Evolution Discovery Platform

The Group Evolution Discovery Platform (GED Platform) was created for the purposes of conducting experiments (Section 7.). The main aim was to implement the GED method and methods by Asur et al. and by Palla et al. Additionally, GED Platform was used to analyze and compare mentioned methods. The scheme of GED Platform, containing all modules, is presented in Figure 5.1.

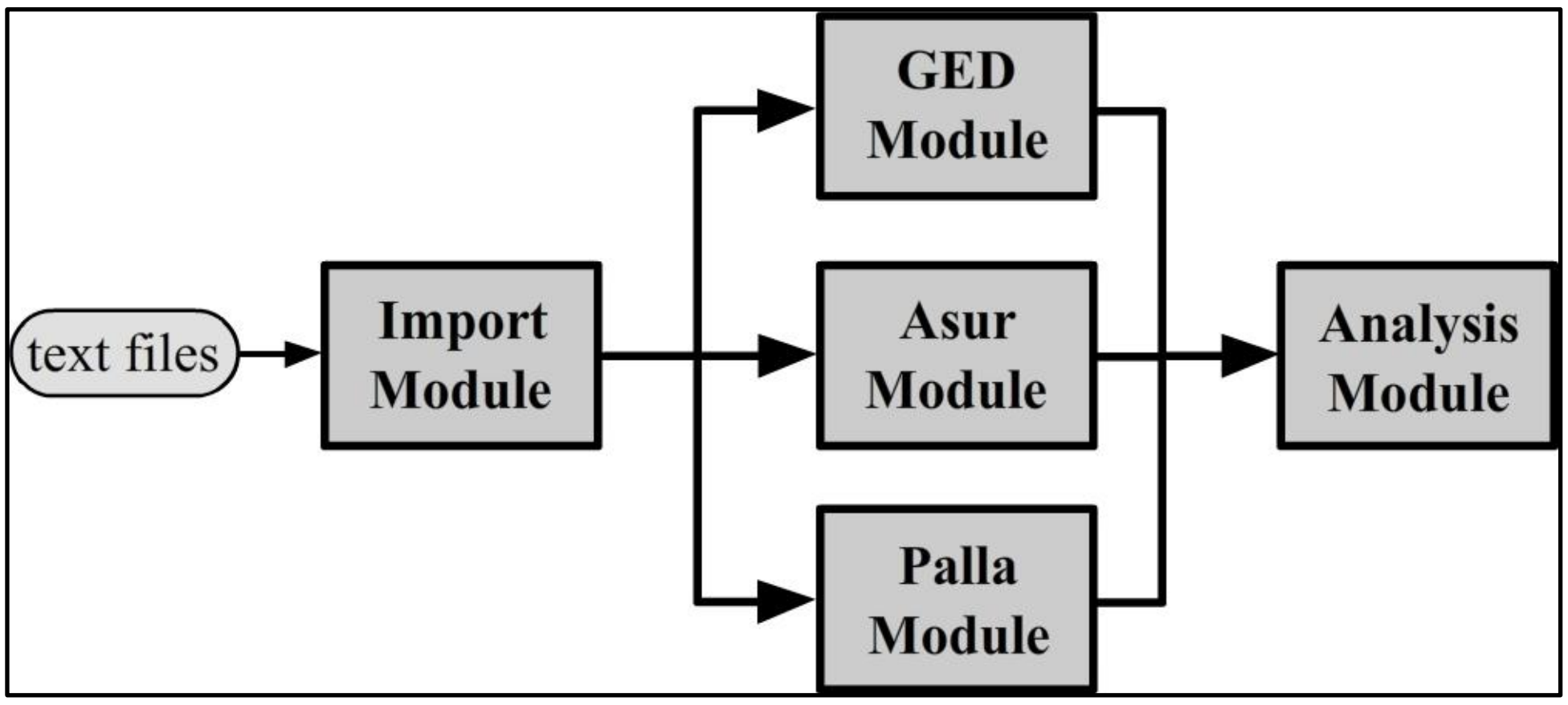

Figure 5.1. Modules in GED Platform.

## 5.1. Data Structures

Each module consists of at least one table. Relations between them are illustrated in Figure 5.2.

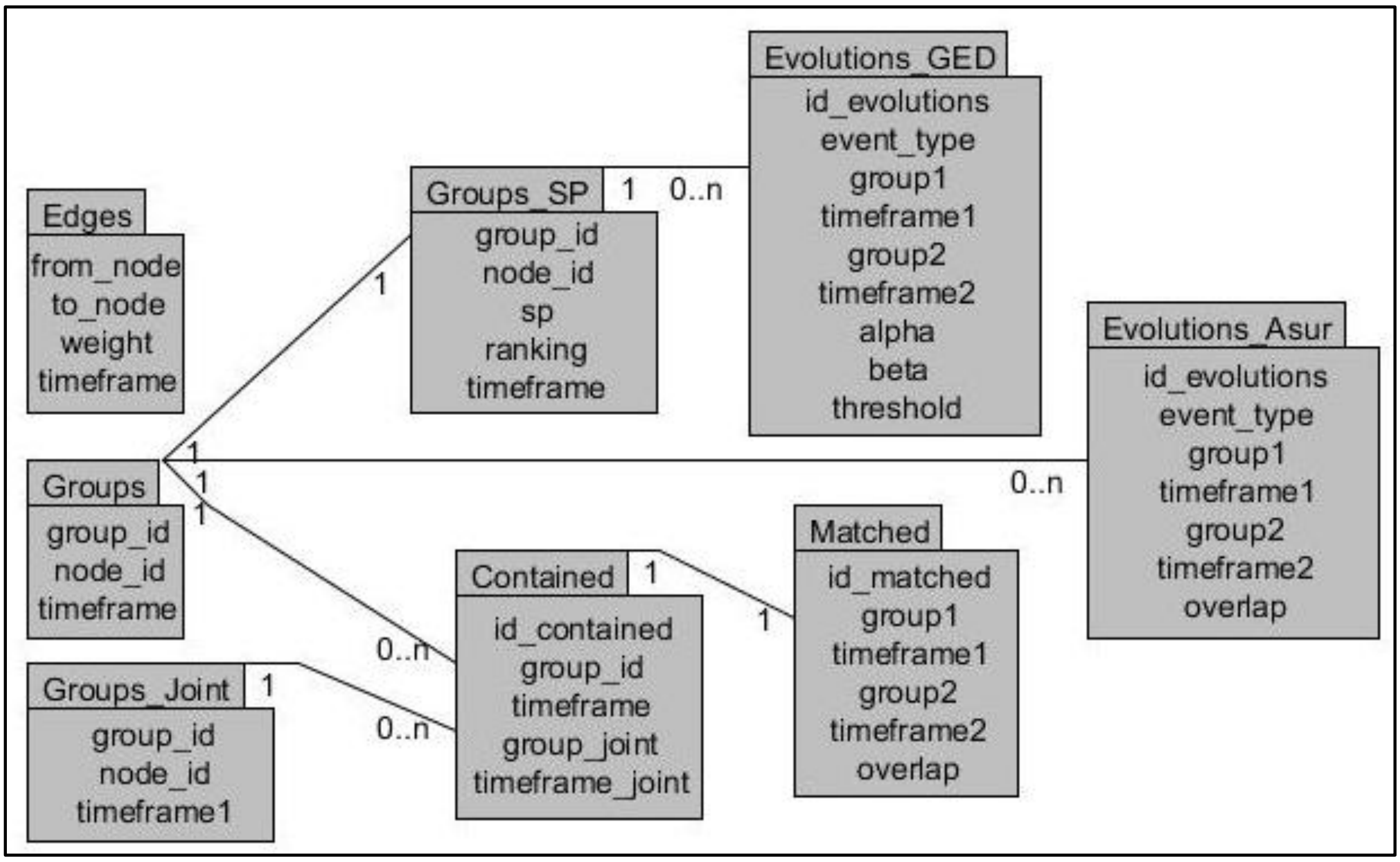

Figure 5.2. Relations between tables within GED Platform.

### *5.1.1. Import Module*

Primarily, import module reads data about groups within social network from text files and stored them in GED Platform's database. User provides tab-delimited files, where each row consists of three columns: group number, member id, timeframe number. All timeframes of a particular social network are stored in one table named Groups, different networks or networks obtained with different grouping methods are stored in separate tables. Scheme of the table is presented in Table 5.1.

| PK | Column name | Data type | Allow nulls |
|---|---|---|---|
| 🔑 | group_id | smallint | No |
| 🔑 | node_id | int | No |
| 🔑 | timeframe | tinyint | No |

Table 5.1. Scheme of table Groups, storing data about groups and members within social network.

For the needs of GED module another data, data about commitment value in the entire network, are imported as well. Each row of tab-delimited files consist of four columns: from node, to node, weight, timeframe. The data for the whole network are stored in the same table named Edges. Structure of the table is presented in Table 5.2.

| PK | Column name | Data type | Allow nulls |
|---|---|---|---|
| 🔑 | from_node | int | No |
| 🔑 | to_node | int | No |
| | weight | float | No |
| | timeframe | tinyint | No |

Table 5.2. Scheme of table Edges, storing data about commitment value within network.

Palla module, in turn, demands data about groups in the joint graph achieved by joining two consecutive timeframes. The data are stored in table Groups_Joint, which structure is identical to table Groups presented in Table 5.1.

### *5.1.2. GED Module*

The main task of GED module is tracking group evolution according to GED method (Section 4.). To make it possible GED module also has to calculate social position (Section 2.4.1.) of a members in all groups from a network. To do so data gathered by import module are used, especially data about groups and members within social network (Table 5.1) and commitment value between members (Table 5.2). Groups with calculated social position are stored in table named Groups_SP, which scheme is presented in Table 5.3.

| PK | Column name | Data type | Allow nulls |
|---|---|---|---|
| 🔑 | group_id | smallint | No |
| 🔑 | node_id | int | No |
| | sp | float | Yes |
| | ranking | int | Yes |
| 🔑 | timeframe | tinyint | No |

Table 5.3. Scheme of table Groups_SP, storing data about groups, members and their social position within social network.

Having data about groups, members and their social position GED module can proceeds its main task – tracking group evolution. By calculating inclusion measures and following GED method's rules new table named Evolutions_GED is filled with the data about groups evolutions. Structure of the table is showed in Table 5.4.

| PK | Column name | Data type | Allow nulls |
|---|---|---|---|
| 🔑 | id_evolutions | int | No |
| | event_type | varchar | No |
| | group1 | int | Yes |
| | timeframe1 | tinyint | Yes |
| | group2 | int | Yes |
| | timeframe2 | tinyint | Yes |
| | alpha | tinyint | Yes |
| | beta | tinyint | Yes |
| | threshold | varchar | No |

Table 5.4. Scheme of table Evolutions_GED, storing data about groups evolutions found with GED method.

### *5.1.3. Asur Module*

The only task of Asur module is discovering group evolution according to Asur et al. method (Section 3.2.1.). Based on the data from Table 5.1 overlapping is calculated and new table named Evolutions_Asur is filled with the data about groups changes. Scheme of the table is presented in Table 5.5.

| PK | Column name | Data type | Allow nulls |
|---|---|---|---|
| 🔑 | id_evolutions | int | No |
| | event_type | varchar | No |
| | group1 | int | Yes |
| | timeframe1 | tinyint | Yes |
| | group2 | int | Yes |
| | timeframe2 | tinyint | Yes |
| | overlap | float | Yes |

Table 5.5. Scheme of table Evolutions_Asur, storing data about groups evolutions discovered with Asur et al. method.

### *5.1.4. Palla Module*

The last module for uncovering communities evolutions is Palla module. Based on the data from tables Groups and Groups_Joint new table named Contained is filled with data about groups from the single timeframes contained in groups from the joint graphs. Structure of the table is showed in Table 5.6. Next, groups located in the same joint graph are matched based on the highest overlap and results are saved in table Matched, which structure is presented in Table 5.7.

| PK | Column name | Data type | Allow nulls |
|---|---|---|---|
| 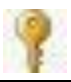 | id_contained | int | No |
| | group_id | int | Yes |
| | timeframe | tinyint | Yes |
| | group_joint | int | Yes |
| | timeframe_joint | tinyint | Yes |

Table 5.6. Scheme of table Contained, storing data about groups from the single timeframes contained in groups from the joint graphs.

| PK | Column name | Data type | Allow nulls |
|---|---|---|---|
| 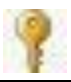 | id_matched | int | No |
| | group1 | int | Yes |
| | timeframe1 | tinyint | Yes |
| | group2 | int | Yes |
| | timeframe2 | tinyint | Yes |
| | overlap | float | Yes |

Table 5.7. Scheme of table Matched, storing data about matched groups.

### *5.1.5. Analysis Module*

The last module in GED Platform is module for analyzing and comparing results. Functions implemented in this module allows to find differences between methods for tracking group evolution. All functions are briefly described in Table 5.8.

| Function name (input) | Description | Tables used |
|---|---|---|
| Inclusion (group1, group2) | Provides detailed information about selected groups, such as: inclusions, sizes, social position of cores, total social positions, intersection. | Groups_SP |
| Migration (group1) | Provides detailed information about single group evolution, such as: size, core size, number of nodes migrated, average ranking of migrated nodes before migration, average ranking of migrated nodes after migration, size of new group. | Groups_SP |
| Compare (table1, table2) | Joins results from two methods for tracking evolution in order to show which events are assigned to each pair of groups by both methods. Also provides events found by one method and omitted by second one. | Evolutions_GED, Evolutions_Asur, Matched_Palla |
| Evolution (group1) | Provides evolution of selected group in next/previous timeframe, i.e. shows which groups are matched with selected group. | Evolutions_GED, Evolutions_Asur, Matched_Palla |
| Evolution (table1) | Shows all possible processes of evolution for all groups from all timeframes. | Evolutions_GED, Evolutions_Asur, Matched_Palla |

Table 5.8. Functions implemented in analysis module.

# 6. Email Communication Data

The email communication data was gained from Wroclaw University of Technology (WrUT), which extracted exchange of messages from its server logs.

## 6.1. Data Description

Data set was delivered in WrUT.zip and it size was 2.7 MB. Unpacked data was divided into fourteen text files (1.txt – 14.txt) and their size was 20.2 MB. All text files contained information about temporal social network (see section 2.5.) which consists from fourteen 90-days timeframes (timestamps from server logs were used to determine exact dates). Timeframes are overlapping with the 45-days overlap, i.e., the first timeframe begins on the 1st day and ends on the 90th day, second begins on the 46th day and ends on the 135th day and so on. The whole data set was collected within period of February 2006 – October 2007 and consists of 5.845 members and 149.344 relations. Each row of the single text file means relation between two members of Wroclaw University of Technology. The relation in this case is exchanging emails and is represented by pair of members and weight of relation between them, Figure 6.1.

```
4376;27588;0,001924927815206929740
4376;28598;0,004812319538017324350
4376;59745;0,000962463907603464870
```

Figure 6.1. Example of the data. WrUT member 1 id (•), WrUT member 2 id (•), weight of relation (•).

Basic information about data:

- Provider – Wroclaw University of Technology
- Size – 20.2 MB (plain text)
- Time period from February 2006 to October 2007
- Number of members – 5.845
- Number of relations – 149.344
- Number of timeframes – 14
- Timeframe interval – 90 days
- Timeframe overlap – 45 days

## 6.2. Data Pre-processing

First, the data had to be converted to fulfil the input requirements of both algorithms for extracting groups. Fortunately both methods, CPM implemented in CFinder [26] and Blondel implemented in a Workbench for Network Scientists (NWB) [49], allowed common data format, tab–delimited text files. Therefore, semicolon was replaced by tab and comma in weight of relation was replaced by point. In order to improve execution time of the algorithms, weight of relation between members was rounded to four decimal places. New format of the data is presented in Figure 6.2.

| | | |
|---|---|---|
| 4376 | 27588 | 0.0019 |
| 4376 | 28598 | 0.0048 |
| 4376 | 59745 | 0.0010 |

Figure 6.2. The data after conversion. WrUT member 1 (•), WrUT member 2 (•), weight of relation (•).

Next, each timeframe of the temporal social network of email communication data was grouped using both methods. Output file of the CFinder software contained the groups sorted in ascending order with assigned members of WrUT, Figure 6.3. The groups may share members, in Figure 6.3 member 615 belongs to group 14 and 16.

```
14: 14151 1154 96 615 1153 5383
15: 2865 853 1225 15866 315 4132
16: 4513 14151 615 1160 6535 5861
```

Figure 6.3. The output of the CFinder software. Group number (•), members of the particular group (•), member 615 (•).

Output file of the NWB application contained the members assigned to the groups on particular hierarchy level. Each member might be a part of only one community at each level of hierarchy, Figure 6.4.

| | | | | |
|---|---|---|---|---|
| 63 | "83" | "community_12" | "community_3" | "community_3" |
| 64 | "292" | "community_12" | "community_3" | "community_3" |
| 65 | "628" | "community_27" | "community_6" | "community_6" |

Figure 6.4. The output of the NWB application. ID (•), member id (•), members communities at particular hierarchy level – from the lowest to the highest (•).

Both output data sets were converted to the same format, tab–delimited text files containing group number, member id, timeframe number, Figure 6.5.

| | | |
|---|---|---|
| 15 | 178 | 7 |
| 15 | 228 | 7 |
| 16 | 292 | 7 |

Figure 6.5. The data after final conversion. Group number (•), member id (•), timeframe number (•).

After final conversion data sets were imported to the common MS SQL database, but to separate tables. Output of the CFinder was imported to the table Groups_CPM, and output of the NWB was imported to the table Groups_Blondel. Both tables had the same structure consisted of three columns: group_id was the type of smallint, node_id was the type of int, timeframe was the type of tinyint. Number of rows in the table Groups_CPM was 29.650, while in the table Groups_Blondel 65.639.

At this stage data sets were ready to run method for tracking evolution provided by Asur et al. The GED method with the simplified version of inclusion measure could be also run at this point of data preparations, however it is recommended to use any measure determining members importance within a group. Therefore, according to [45] an algorithm for calculating social position (SP) was implemented in T-SQL language and run on data from the tables Groups_CPM and Groups_Blondel. As a result new tables were created, Groups_CPM_SP and Groups_Blondel_SP, with additional column of float type for SP value.

Method by Palla et al. required more preparations, because each pair of following timeframes had to be merged into single networks (graphs) and extracted again by CFinder. When grouping was finished output files were converted to format presented in Figure 6.5, but in this case timeframe number referred to the first timeframe in the pair. Afterwards, data set was imported to the database to the table Groups_CPM_Joint, which had identical structure as the table Groups_CPM. Number of rows in this table was 36.153. All tables in the database necessary to conduct experiments are presented in Table 6.1.

| Table name | Description | Size [MB] | No. of rows |
|---|---|---|---|
| **Groups_CPM** | List of members assigned to groups in each timeframe, extracted with CPM | 0,695 | 29.650 |
| **Groups_Blondel** | List of members assigned to groups in each timeframe, extracted with Blondel | 1,016 | 65.639 |
| **Groups_CPM_SP** | List of members and their social position, assigned to groups in each timeframe, extracted with CPM | 1,188 | 29.650 |
| **Groups_Blondel_SP** | List of members and their social position, assigned to groups in each timeframe, extracted with Blondel | 1,656 | 65.639 |
| **Groups_CPM_Joint** | List of members assigned to groups in networks obtained by joining two following timeframes, extracted with CPM | 0,883 | 36.153 |

Table 6.1. Basic information about tables required for experiments.

# 7. Experiments

The main aim of the experiments was to investigate the features of GED method, such as accuracy, flexibility, execution time, etc. Accuracy of a method is the ability to catch evolutions, i.e. how many pairs of groups from different timeframes which share nodes can be found by the algorithm. Flexibility, in turn, determines how much influence on a results has user by adjusting methods parameters. Moreover, the influence of thresholds values on the results was examined. Differences in results obtained by GED with different user importance measures were also investigated. Lastly, GED method and methods by Asur et al. and by Palla et al. were compared on groups obtained with CPM algorithm and Blondel algorithm. The study was focused on accuracy, flexibility and execution time of particular methods, but other aspects, e.g. ease of implementation, design, were also took into account.

## 7.1. Test Environment

The first step of the experiments, extraction of the communities by grouping algorithms, was conducted on stationary computer with computational power 3 GHz (Intel Pentium Dual Core) and 2 GB of RAM memory, the operating system was Microsoft Windows 7 Professional. Both software packages, CFinder and NWB, required Sun's Java Runtime Environment (JRE) in version not lower than 1.4. All single timeframes were extracted without any problems, but extraction of networks obtained by joining two consequence timeframes was unsuccessful because of the size and density of those networks. Therefore, another computer, with computational power 2,8 GHz (Intel Pentium Core 2 Duo) and 8 GB of RAM memory, was used to extract mentioned networks, this time with success.

For the next step, implementing methods for group evolution discovery, another computer, with computational power 1,7 GHz (Intel Pentium Core Duo) and 2 GB of RAM memory, was used. The platform used for implementing methods was Microsoft SQL Server Management Studio 2005. A list of the hardware and software utilized in experiments together with tasks performed on each computer is presented in Table 7.1.

| Id | Computational power | Software needed | Tasks |
|---|---|---|---|
| 1. | Intel Pentium Dual Core 3 GHz, 2 GB of RAM. | CFinder, NWB. | Extracting communities from each timeframes. |
| 2. | Intel Pentium Core 2 Duo 2,8 GHz, 8 GB of RAM. | CFinder. | Extracting communities from networks obtained by joining two consecutive timeframes. |
| 3. | Intel Pentium Core Duo 1,7 GHz, 2 GB of RAM. | MS SQL 2005. | Data set conversions (pre-processing), implementing and running methods for tracking community evolution. |

Table 7.1. A list of hardware and software used in experiments.

## 7.2. Experiment Based on Overlapping Groups Extracted by CPM

In the first experiment, as a method for group extraction, CPM implemented in CFinder (www.http://cfinder.org/) was used. The groups were discovered for *k*=6 and for the directed and unweighted social network. CFinder extracted from 80 to 136 groups for the timeframe (avg. 112 per timeframe, Table 7.2). The time needed to extract single time window on computer 1. from Table 7.1 varied from 1minute to 20 hours, depending on the size and density of the network. The average size of the group was 19 nodes (Table 7.2), the smallest groups had size 6, because of *k* parameter, and the biggest one was 613 in time window 10.

| **Time window** | **No of nodes in network** | **Number of groups** | **Avg size of a group** |
|---|---|---|---|
| 1 | 1585 | 136 | 15,7 |
| 2 | 1616 | 128 | 17,2 |
| 3 | 1579 | 129 | 16,8 |
| 4 | 1131 | 82 | 17,4 |
| 5 | 1067 | 105 | 13,9 |
| 6 | 1867 | 96 | 23,6 |
| 7 | 2056 | 117 | 23 |
| 8 | 1999 | 123 | 21,6 |
| 9 | 2351 | 108 | 26,8 |
| 10 | 2323 | 119 | 24,8 |
| 11 | 2139 | 98 | 27 |
| 12 | 1557 | 125 | 16,9 |
| 13 | 955 | 121 | 11,2 |
| 14 | 536 | 80 | 9,3 |
| **Avg** | **1626** | **112** | **19** |

Table 7.2. Results of CPM method extraction.

### *7.2.1. GED Method*

As already mentioned, the GED method was implemented in T-SQL language. The method was run frequently with different value of $\alpha$ and $\beta$ thresholds to analyse the influence of these parameters on the method, the results are presented in Table 7.3. The time needed for single run was about 6 minutes. The lowest checked value for the thresholds was set to 50%, which guarantee that at least half of the considered group was contained in the matched group. The highest possible value was of course 100% and means that the studied group is identical with the matched group. The thresholds for *forming* and *dissolving* event was set to 10% based on average group size and intuition.

Average group size is 19 and average core size in this case is 8. Social position of this core is 11 what means that, with threshold 10% for forming (dissolving), groups cannot have more than 4-5 strong nodes (core nodes) or 6-7 weak nodes (nodes outside the core) existing together in one group from previous (next) timeframe in order to assign forming (dissolving) event.

$\overbrace{\frac{5}{19}}^{core} * \overbrace{\frac{7}{19}}^{sp} \approx 9{,}7\%$ – calculations for strong nodes (core nodes). Event assigned, inclusion < 10%.

$\overbrace{\frac{7}{19}}^{mixed} * \overbrace{\frac{6}{19}}^{sp} \approx 12\%$ – calculations for few strong and few weak nodes. Event not assigned, inclusion > 10%.

$\overbrace{\frac{7}{19}}^{weak} * \overbrace{\frac{4}{19}}^{sp} \approx 8\%$ – calculations for weak nodes (nodes outside the core). Event assigned, inclusion < 10%.

| Threshold | | Number of | | | | | | | |
|---|---|---|---|---|---|---|---|---|---|
| **α** | **β** | **form** | **dissolve** | **shrink** | **growth** | **continue** | **split** | **merge** | **total** |
| 50 | 50 | 122 | 186 | 204 | 180 | 127 | 517 | 398 | 1734 |
| 50 | 60 | 122 | 186 | 204 | 173 | 124 | 464 | 405 | 1678 |
| 50 | 70 | 122 | 186 | 202 | 157 | 124 | 400 | 421 | 1612 |
| 50 | 80 | 122 | 186 | 203 | 149 | 122 | 311 | 429 | 1522 |
| 50 | 90 | 122 | 186 | 199 | 154 | 122 | 279 | 424 | 1486 |
| 50 | 100 | 122 | 186 | 199 | 156 | 122 | 261 | 422 | 1468 |
| 60 | 50 | 122 | 186 | 190 | 177 | 124 | 531 | 359 | 1689 |
| 60 | 60 | 122 | 186 | 191 | 170 | 120 | 475 | 366 | 1630 |
| 60 | 70 | 122 | 186 | 187 | 152 | 119 | 409 | 384 | 1559 |
| 60 | 80 | 122 | 186 | 187 | 144 | 117 | 314 | 392 | 1462 |
| 60 | 90 | 122 | 186 | 181 | 148 | 117 | 277 | 388 | 1419 |
| 60 | 100 | 122 | 186 | 179 | 149 | 117 | 259 | 387 | 1399 |
| 70 | 50 | 122 | 186 | 179 | 176 | 123 | 543 | 284 | 1613 |
| 70 | 60 | 122 | 186 | 180 | 170 | 119 | 486 | 286 | 1549 |
| 70 | 70 | 122 | 186 | 177 | 156 | 113 | 418 | 298 | 1470 |
| 70 | 80 | 122 | 186 | 174 | 149 | 111 | 317 | 305 | 1364 |
| 70 | 90 | 122 | 186 | 165 | 150 | 111 | 277 | 304 | 1315 |
| 70 | 100 | 122 | 186 | 161 | 152 | 111 | 259 | 302 | 1293 |
| 80 | 50 | 122 | 186 | 172 | 169 | 120 | 553 | 233 | 1555 |
| 80 | 60 | 122 | 186 | 173 | 154 | 117 | 495 | 235 | 1482 |
| 80 | 70 | 122 | 186 | 170 | 137 | 111 | 426 | 244 | 1396 |
| 80 | 80 | 122 | 186 | 165 | 127 | 97 | 324 | 251 | 1272 |
| 80 | 90 | 122 | 186 | 157 | 128 | 96 | 276 | 250 | 1215 |
| 80 | 100 | 122 | 186 | 152 | 129 | 96 | 257 | 249 | 1191 |
| 90 | 50 | 122 | 186 | 172 | 169 | 120 | 553 | 199 | 1521 |
| 90 | 60 | 122 | 186 | 174 | 152 | 117 | 494 | 198 | 1443 |
| 90 | 70 | 122 | 186 | 171 | 132 | 111 | 425 | 199 | 1346 |
| 90 | 80 | 122 | 186 | 165 | 121 | 96 | 324 | 203 | 1217 |
| 90 | 90 | 122 | 186 | 154 | 123 | 91 | 276 | 199 | 1151 |
| 90 | 100 | 122 | 186 | 148 | 123 | 91 | 257 | 199 | 1126 |
| 100 | 50 | 122 | 186 | 176 | 167 | 120 | 549 | 185 | 1505 |
| 100 | 60 | 122 | 186 | 177 | 149 | 117 | 491 | 183 | 1425 |
| 100 | 70 | 122 | 186 | 173 | 127 | 111 | 423 | 180 | 1322 |
| 100 | 80 | 122 | 186 | 166 | 116 | 96 | 323 | 179 | 1188 |
| 100 | 90 | 122 | 186 | 154 | 117 | 91 | 276 | 173 | 1119 |
| 100 | 100 | 122 | 186 | 148 | 115 | 90 | 257 | 173 | 1091 |

Table 7.3. The results of GED computation on overlapping groups extracted by CPM.

Analogously, small groups of size 6 and average core size of 2.5 with social position 3, can contain maximum 2 weak nodes existing in the same group from previous (next) timeframe to be treated as new born (dissolved).

$$\frac{\overbrace{2}^{core}}{6} * \frac{\overbrace{3}^{sp}}{6} \approx 17\%$$

– calculations for two strong nodes (core nodes). Event not assigned, inclusion > 10%.

$\overbrace{\frac{2}{6}}^{\text{mixed}} * \overbrace{\frac{2}{6}}^{\text{sp}} \approx 11\%$ – calculations for one strong node and one weak node. Event not assigned, inclusion > 10%.

$\overbrace{\frac{2}{6}}^{\text{weak}} * \overbrace{\frac{1}{4}}^{\text{sp}} \approx 8\%$ – calculations for two weak nodes (nodes outside the core). Event assigned, inclusion < 10%.

While analysing Table 7.3, it can be observed that with the increase of $\alpha$ and $\beta$ thresholds, the total number of events is decreasing, when $\alpha$ and $\beta$ equals 50% the number is 1734, and with thresholds equal 100% the number is only 1091. This means that the parameters $\alpha$ and $\beta$ can be used to filtering results, preserving events where groups are highly overlapped. Another advantage of having parameters is possibility to adjust the results to ones needs. Setting appropriate levels for $\alpha$ and $\beta$ thresholds effects in tendency to assign particular event more often than another one. For example group $G_{40}$ from timeframe $T_6$ match group $G_{68}$ from timeframe $T_7$ with inclusions equals respectively $I_1(G_{40}, G_{68})$=82% and $I_2(G_{68}, G_{40})$=76%. Then, event assigned between considered groups depends on the value of $\alpha$ and $\beta$, Table 7.4.

| **Threshold** | | **Group1** | **Time window1** | **Event type** | **Group2** | **Time window2** | $I_1$ | $I_2$ |
|---|---|---|---|---|---|---|---|---|
| $\alpha$ | $\beta$ | | | | | | | |
| 70 | 50 | 40 | 6 | growing | 68 | 7 | 82% | 76% |
| 70 | 60 | 40 | 6 | growing | 68 | 7 | 82% | 76% |
| 70 | 70 | 40 | 6 | growing | 68 | 7 | 82% | 76% |
| 70 | 80 | 40 | 6 | merging | 68 | 7 | 82% | 76% |
| 70 | 90 | 40 | 6 | merging | 68 | 7 | 82% | 76% |
| 70 | 100 | 40 | 6 | merging | 68 | 7 | 82% | 76% |
| 100 | 50 | 40 | 6 | merging | 68 | 7 | 82% | 76% |
| 100 | 60 | 40 | 6 | merging | 68 | 7 | 82% | 76% |
| 100 | 70 | 40 | 6 | merging | 68 | 7 | 82% | 76% |
| 100 | 80 | 40 | 6 | no event | 68 | 7 | 82% | 76% |
| 100 | 90 | 40 | 6 | no event | 68 | 7 | 82% | 76% |
| 100 | 100 | 40 | 6 | no event | 68 | 7 | 82% | 76% |

Table 7.4. Influence of thresholds values on assigning event type.

The linear increase of threshold $\alpha$ causes close to linear reduction in number of merging events and slightly decrease of the number of growing events, Figure 7.1 and Figure 7.2.

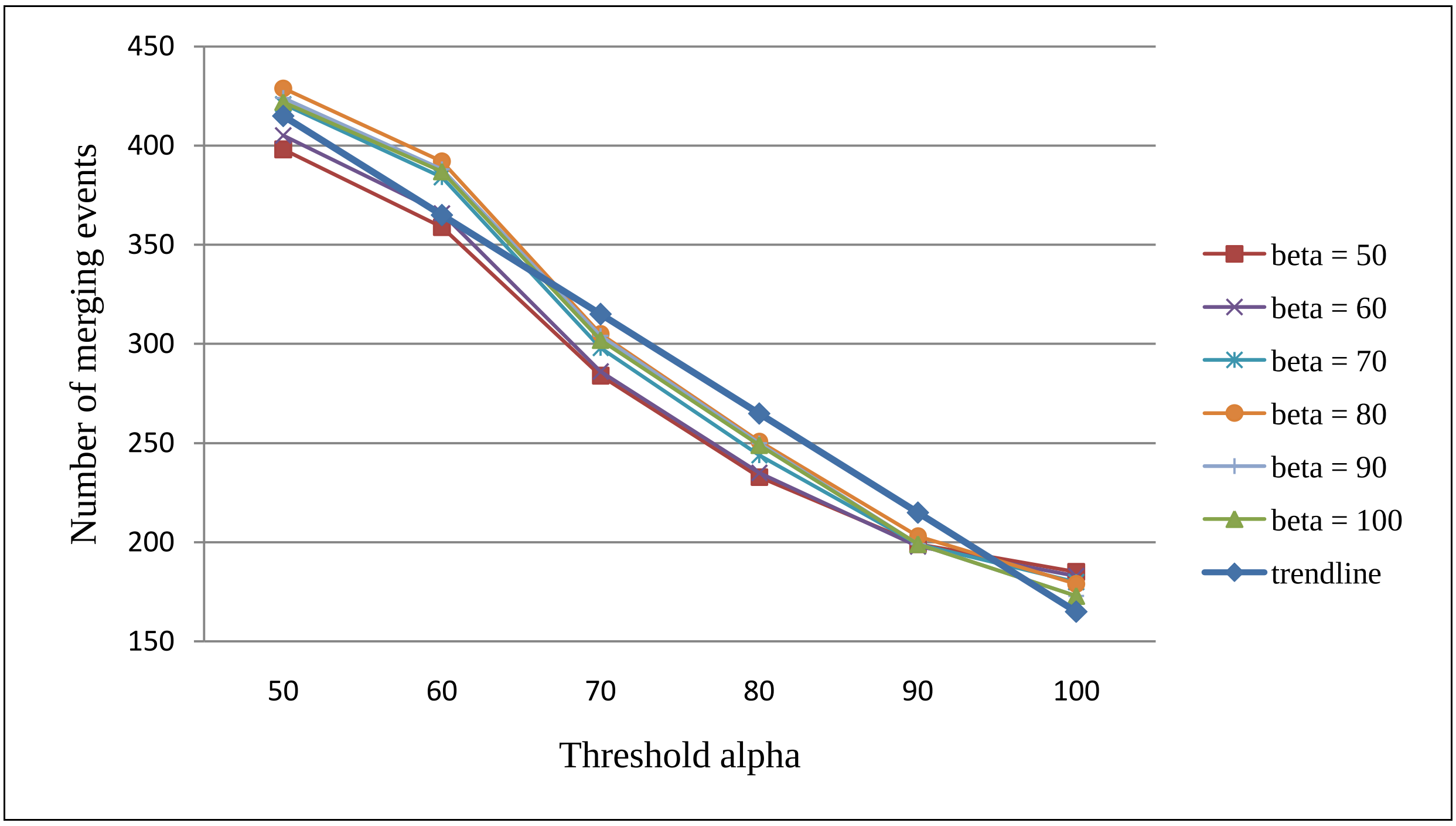

Figure 7.1. Number of merging events for different values of alpha and beta thresholds.

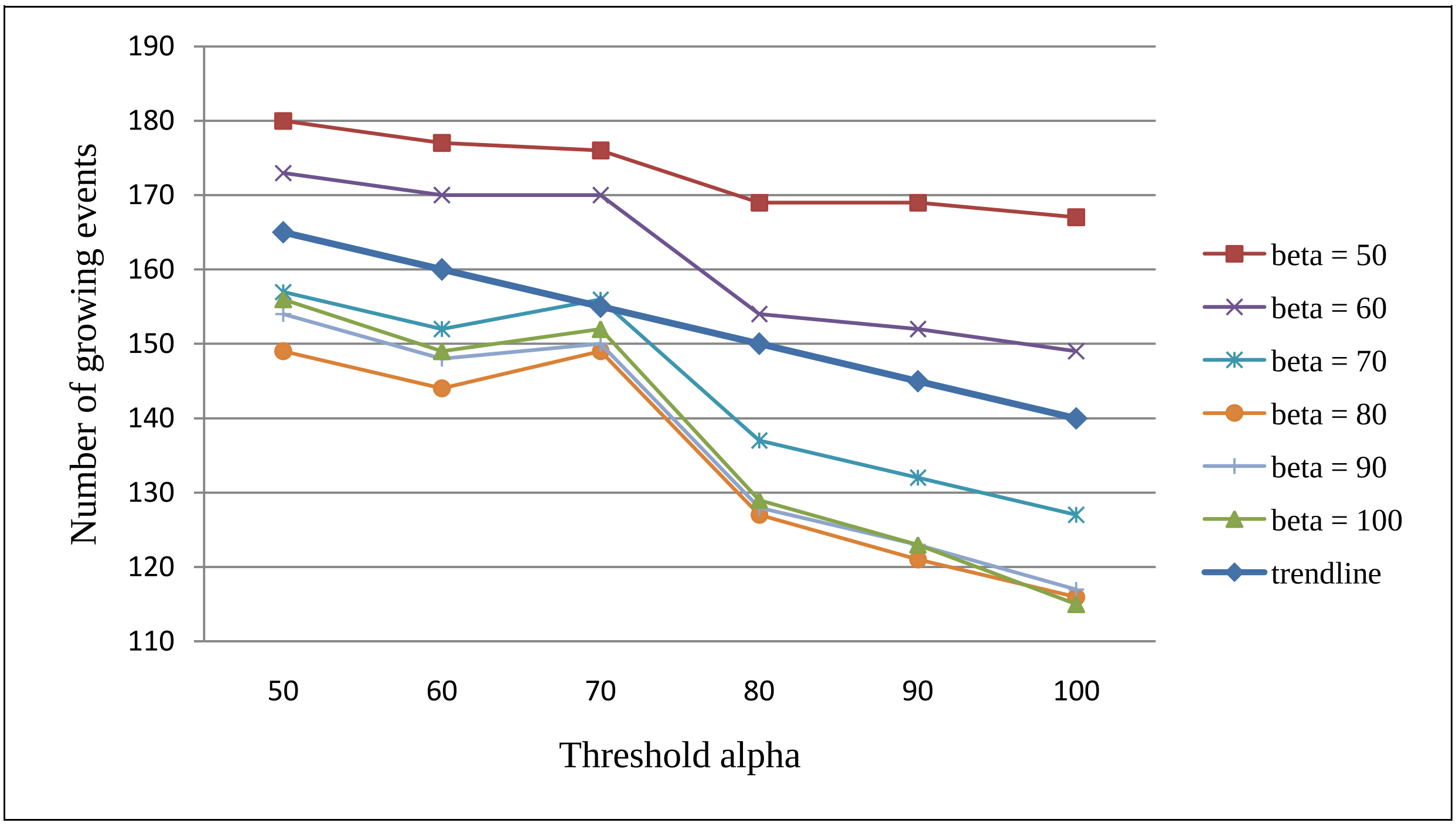

Figure 7.2. Number of growing events for different values of alpha and beta thresholds.

In contrast, with linear increase of threshold $\beta$, the number of splitting events decrease in almost linear way and the number of shrinking events slightly decrease as well, Figure 7.3 and Figure 7.4.

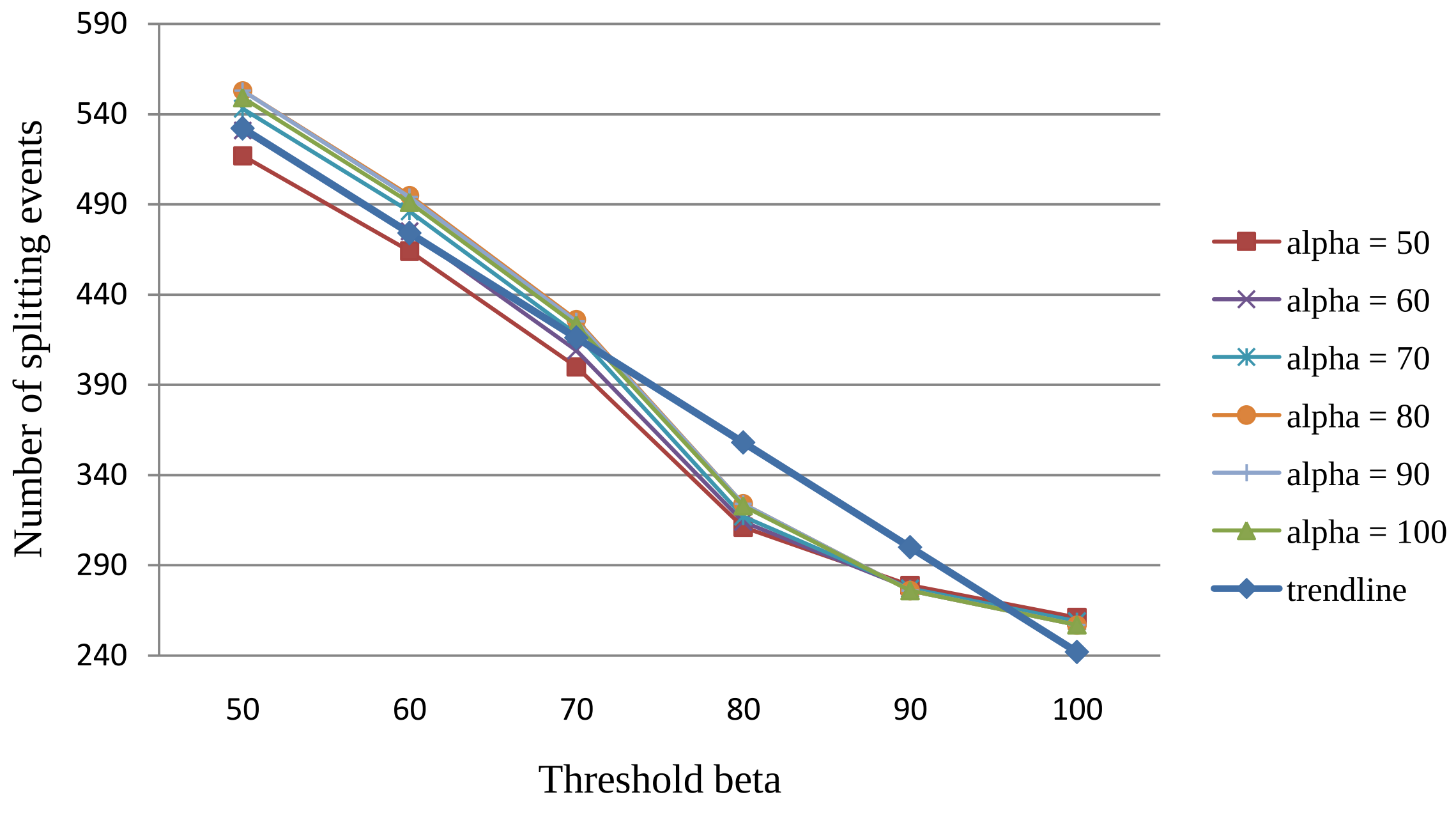

Figure 7.3. Number of splitting events for different values of beta and alpha thresholds.

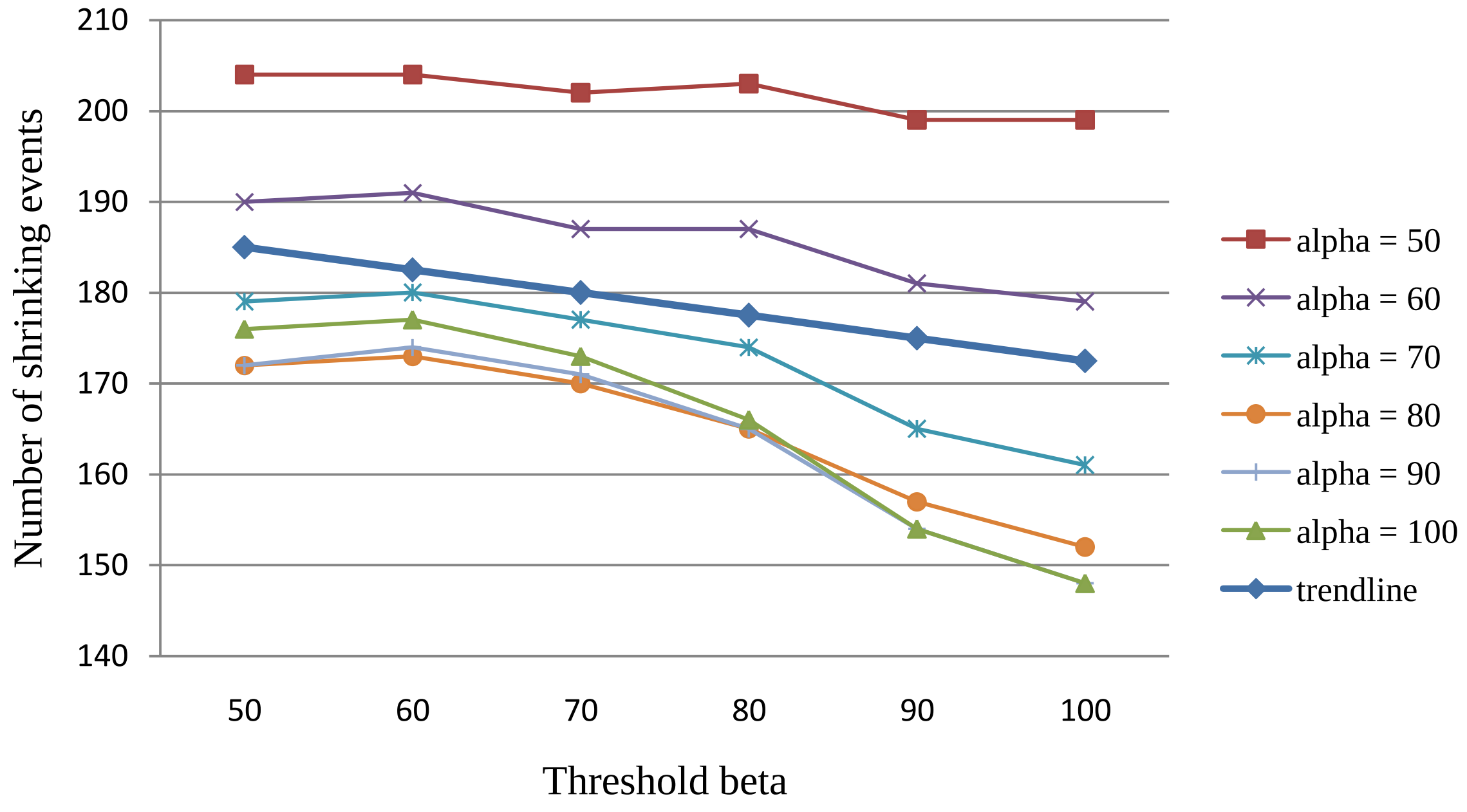

Figure 7.4. Number of shrinking events for different values of beta and alpha thresholds.

It is a consequence of the algorithm structure, raising the thresholds makes it harder to match the groups (see Figure 4.3). Furthermore, dissolving event occurs more often than forming event. The main reason is the fact that the last timeframe covers only the period of summer holidays, and as a result the email exchange is very low. This causes the groups to be small and have low density.

GED method found 90 continue events when both inclusions of groups are equal to 100%.

### *7.2.2. Method by Asur et al.*

The method by Asur et al. was implemented in T-SQL language as well. The authors suggested to set 30% or 50% as an overlapping threshold for merge and split. In experiment threshold was set to 50%. It took more than 5.5 hours to calculate events between groups in all fourteen time windows. The total number of events found by Asur et al. method is 1526, from which 90 are continuation, 18 are forming, 29 are dissolving, 703 are merging and 686 are splitting.

Such a small number of continuing events is caused by very rigorous condition, which requires for groups to remain unchanged. Small amount of forming (dissolving) events came from another strong condition, which state that none of the nodes from the considered group can exist in network at previous (following) time window. A huge number of merging (splitting) events is a result of low overlapping threshold for merge (split).

However, it has to be noticed that these numbers are slightly overestimated. Method by Asur et al. allows that one pair of groups has assigned more than one type of event. This leads to anomalies presented in Figure 7.5a and in Table 7.5.

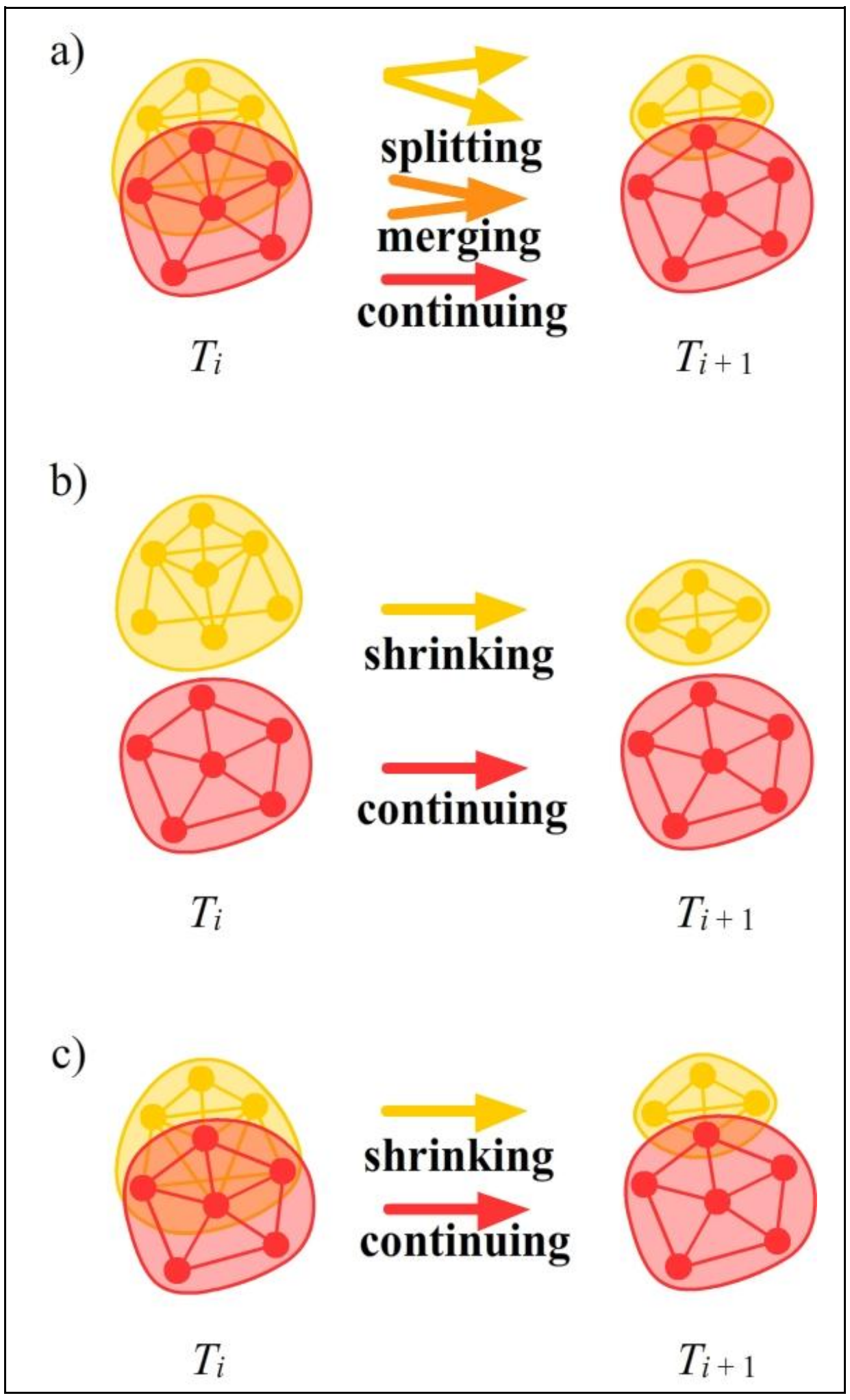

Figure 7.5. Evolution of two overlapping groups. a) anomalies generated by Asur et al. method, b) simplified case, when groups are disjoint c) events assigned correctly by GED method.

Community coloured red overlaps with community coloured yellow in timeframe $T_i$ by sharing four members. In next timeframe $T_{i+1}$ community marked with red has exactly the same structure, while yellow community brake up relations with three members from shared area. If these groups were disjoint, the events would be assigned without any doubt (Figure 7.5b). But they are not, and it is up to matching algorithm to deal with this situation. Intuition says that red community continue its existence, while yellow shrinks or splits into two groups, both events are correct (but not both at the same time). According to method by Asur et al. community coloured red is continuing and also merging with yellow one. In the meantime community marked with yellow splits into communities marked with yellow and red, and also merges with red community (Figure 7.5a and Table 7.5). Such a case should not appear when condition for continuing event in Asur et al. method is so rigorous. What is more, Asur et al. defined merging event as a joining members (implicitly different members) from two groups into single one, and in the situation presented above joining members are shared by both groups, so there should be no merging event at all. Possible explanation for anomalies is that method by Asur et al. is not designed for overlapping communities. Further experiments with disjoint groups should clarify this assumption (section 7.3.2.).

| **Group1** | **Time window1** | **Event type** | **Group2** | **Time window2** | **Overlap** |
|---|---|---|---|---|---|
| 13• | 1• | splitting | 2• | 2• | 57% |
| 13• | 1• | splitting | 9• | 2• | 57% |
| 13• | 1• | merging | 9• | 2• | 57% |
| 1• | 1• | merging | 9• | 2• | 100% |
| 1• | 1• | continuing | 9• | 2• | 100% |

Table 7.5. Anomalies generated by Asur et al. method.
Colours marks groups illustrated in Figure 7.5.

The total number of anomalies is 128 cases, 8% of all results. More than a half of these cases are groups with split and merge event into another group at the same time. The rest of the cases are even worse, because one group has continue and split or merge event into another group simultaneously (Figure 7.5a, Table 7.5). Therefore the total number of "distinct" events found by Asur et al. is 1398.

A great advantage of the method is ease of implementation. Asur et al. provided simple formulas, together with pseudo-code (section 3.2.1.), which can be implemented in any programming language. On the other hand, a big disadvantage of the method is very low flexibility. Only threshold for splitting/merging can be adjusted, all other thresholds are constant.

### *7.2.3. GED Method vs. Asur et al. Method*

As already mentioned, the computation time for Asur et al. method was more than 5.5 hours, while for GED it took less than 4 hours to calculate whole Table 7.3. The single run of GED method lasted less than 6 minutes, so it is over 50 times faster than method by Asur et al.

The GED method run with thresholds equals 50% found 721 events which method by Asur et al. has not. Such a big lack in results obtained with Asur et al. method is caused mostly by rigorous conditions for assigning events and almost no flexibility of the method. From the other hand Asur et al. method found 399 events which GED method run with thresholds 50% has not. However, it is not treated as a lack in GED's results because all these events have both inclusions below 50%, therefore GED algorithm skips them in purpose (because of thresholds value). To prove this, GED method was run with thresholds equal 10% and this time none of events found by Asur et al. method were skipped by GED method.

Furthermore, Asur et al. did not introduced shrinking and growing events, which effects in assigning splitting and merging events or, in the worst case, missing the event, Figure 7.6. When two groups in successive time windows differ only by one node they will not be treated as continuation (since the overlapping is below 100%) and might not be treated as merging (splitting) if there does not exist another group fulfilling the requirements for merging (splitting). Such a case is not possible in GED method, which through the change of inclusion thresholds allows to adjust the results to user's needs.

As demonstrated in Figure 7.6a, the community at timeframe $T_{i+1}$ has three members, who not belongs to the group at timeframe $T_i$. Additionally, the members are not present at timeframe $T_i$ in any other group what, results in omitting the event by Asur et al. method despite the fact that overlapping is 63%. In case when mentioned members are present at timeframe $T_i$ in any group (Figure 7.6b) the method assigns merging event, which in this case is not perfect but still better than nothing. In the situation presented in Figure 7.6a GED method assigns growing event (Figure 7.6c) and in the second considered case GED decides based on the size of the groups at timeframe $T_i$ and social position of their members in the community at the following timeframe $T_{i+1}$. In this case GED assigned growing event to bigger group and merging event to smaller one (Figure 7.6d), however if the groups would be more equal or core of the group at timeframe $T_{i+1}$ would come from the pink group (smaller one) then GED method would assign merging event to both groups at timeframe $T_i$.

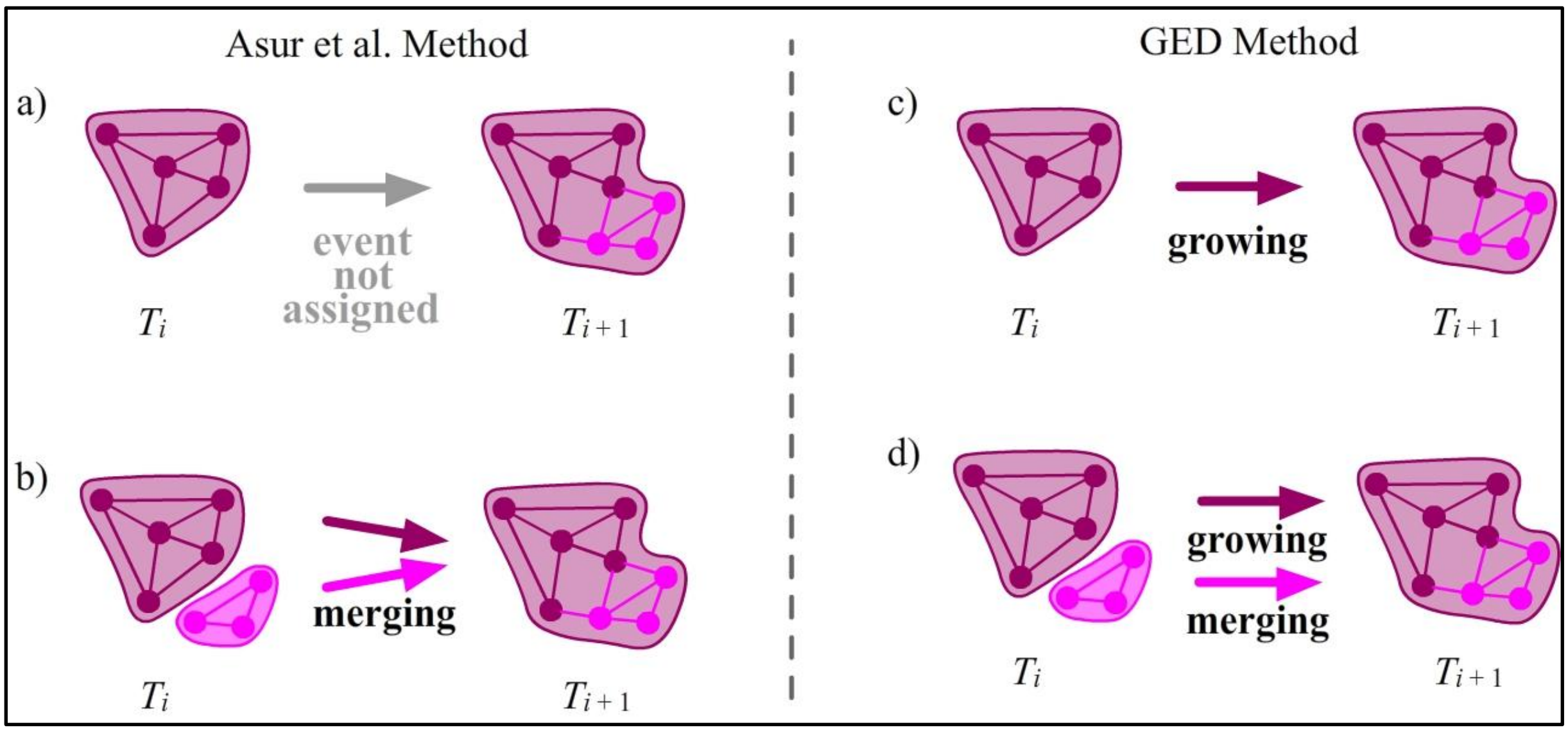

Figure 7.6. Lack of the growing event in method by Asur et al.

Lastly, GED method is free of anomalies occurs in Asur et al. method. While algorithm by Asur et al. is assigning multiple events between two groups (Figure 7.5a), GED method assigns correct events (Figure 7.5c). In the exampled picture method assigned shrinking event between yellow communities because core of the group (the most important members) stayed in the group. If the members who left the group had higher social position than rest of the group, then algorithm would assigned splitting event, which is also correct in this situation.

In most cases events assigned by both methods are the same, however stated analysis proves that GED method is not only faster but also more accurate and much more flexible than method by Asur et al.

### *7.2.4. Method by Palla et al.*

The method delivered by Palla et al. was also implemented in T-SQL language, but it required more preparations with data set than method by Asur et al. Apart from extracting groups in all time windows, another group extraction was needed. The data from two consecutive timeframes were merged into single graphs, from which groups were extracted with CPM method. As easy to count, group extraction had to be calculated additional thirteen times, some of them took only five minutes to calculate, but there were also those that lasted up to two days.

Palla et al. designed method in order to catch all matched pairs of groups, even if they overlap in the slightest way, sharing only one node. The great advantage of this approach is that no event will be ignored. However, if one takes into account the fact, that Palla et al. only showed which event types may occur (and did not provide the algorithm to assign them), analysis of the group evolution during its life is very difficult and cumbersome. Each case of assigning event must be considered individually and with a huge number of possibilities it is very hard to find key match. On average one group from joint graph contains five groups from single timeframes, what gives not less than three and up to six possibilities of matching the groups. Data in Table 7.6 presents groups extracted from the single timeframes $T_8$ and $T_9$ contained in the group $G_{19}$ extracted from the joint graph $T_{8,9}$. Three of the groups are from timeframe $T_8$ and another three from $T_9$, therefore groups can be matched on nine different ways, Table 7.7.

| **Group id** | **Time window** | **Joins id** | **Time windows** |
|---|---|---|---|
| 68 | 8 | 19 | 8-9 |
| 83 | 8 | 19 | 8-9 |
| 102 | 8 | 19 | 8-9 |
| 23 | 9 | 19 | 8-9 |
| 26 | 9 | 19 | 8-9 |
| 49 | 9 | 19 | 8-9 |

Table 7.6. Groups from the single timeframes $T_8$ and $T_9$ contained in the group $G_{19}$ extracted from the joint graph $T_{8,9}$.

| **Group1** | **Time window1** | **Group2** | **Time window2** | **Overlap** |
|---|---|---|---|---|
| 68 | 8 | 23 | 9 | 13% |
| 68 | 8 | 26 | 9 | 8% |
| 68 | 8 | 49 | 9 | **53%** |
| 83 | 8 | 23 | 9 | 6% |
| 83 | 8 | 26 | 9 | **69%** |
| 83 | 8 | 49 | 9 | 8% |
| 102 | 8 | 23 | 9 | 0% |
| 102 | 8 | 26 | 9 | **50%** |
| 102 | 8 | 49 | 9 | 4% |

Table 7.7. All possibilities of matching groups presented in Table 7.6. In bold the highest overlapping for each group from time window1.

Then pairs can be sorted based on overlap in descending order and only the highest overlap for each group is taken. In Table 7.8 green colour marks final matching, each group from timeframe $T_8$ has assigned one group in $T_9$. Colour red marks the group from timeframe $T_9$ which is not assigned to any of the groups from previous timeframe. Authors of the method did not describe how to treat groups like this. Moreover, Palla et al. in did not explain how to choose best match for the community, which in the next timeframe has the highest overlapping value with two different groups. The authors only defined case when there is single highest overlapping for each group.

| Group1 | Time window1 | Group2 | Time window2 | Overlap |
|---|---|---|---|---|
| 83 | 8 | 26 | 9 | **69%** |
| 68 | 8 | 49 | 9 | **53%** |
| 102 | 8 | 26 | 9 | **50%** |
| 68 | 8 | 23 | 9 | 13% |
| 83 | 8 | 49 | 9 | 8% |
| 68 | 8 | 26 | 9 | 8% |
| 83 | 8 | 23 | 9 | 6% |
| 102 | 8 | 49 | 9 | 4% |
| 102 | 8 | 23 | 9 | 0% |

Table 7.8. Final matching of groups based on the highest overlap. Groups marked with green are matched, $G_{23}$ marked with red is not assigned to any group from previous timeframe.

The total number of matched pairs found by Palla et al. method is 9797, from which 4183 pairs have overlap higher than 0%. The authors did not specify how to interpret the groups matched with overlap equal 0%, but intuition suggests to omit these records, since they do not share any nodes. There are 90 cases when matched pairs have overlap equal 100%, which corresponds to continuation event in Asur et al. method.

### *7.2.5. GED Method vs. Palla et al. Method*

As noted before, the method by Palla et al. needed additional preparations to run the experiment, which lasted almost week, therefore GED method, despite the fact that it requires calculated social position, is incomparable faster.

Great advantage of method by Palla et al. is catching all matched pairs of groups. As in case when comparing GED method with algorithm by Asur et al., Palla et al. method found more matched pairs than GED method run with thresholds equal 50%. Again, it is not treated as a lack in GED's results since all these events have both inclusions below 50%, Table 7.9. To confirm that, results obtained with GED on thresholds equal 10% were compared, and this time all matched pairs found by Palla et al. method have both inclusions below 10%. What is more, GED method found 308 events which method by Palla et al. has not. These events are forming and dissolving.

| Group1 | Time window1 | Group2 | Time window2 | Overlap | $I_1$ | $I_2$ |
|---|---|---|---|---|---|---|
| 65 | 1 | 115 | 2 | 45% | 40% | 25% |
| 44 | 6 | 78 | 7 | 40% | 19% | 46% |
| 78 | 12 | 94 | 13 | 31% | 13% | 27% |
| 81 | 9 | 54 | 10 | 20% | 11% | 7% |
| 72 | 10 | 96 | 11 | 13% | 2% | 11% |
| 117 | 3 | 23 | 4 | 7% | 11% | 0% |
| 91 | 9 | 3 | 10 | 1% | 10% | 0% |

Table 7.9. Events found with method by Palla et al. omitted by GED method because of low inclusions values.

Another problem with Palla et al. method is lack of algorithm for assigning events. It is very difficult and time consuming to determine event for the group in the next timeframe, not to mention all fourteen, Table 7.6, Table 7.7, Table 7.8. So, the GED method with its fully automatic algorithm for assigning events is much more useful and convenient.

Summing up, GED method is beyond compare when it comes to execution time, it is also definitely more specific in assigning events and therefore much more effective for tracking group evolution. Method by Palla et al. was helpful only to check if GED method found all events between groups.

Additionally, Palla et al. method requires usage of CPM method which is big disadvantage because it cannot be utilized with other community extraction method, while GED method may be applied for any existing group extraction algorithms.

## 7.3. Experiment Based on Disjoint Groups Extracted by *Blondel*

In the second experiment Blondel et al. method was used for community detection. The algorithm is implemented for example in a Workbench for Network Scientists (NWB). The groups were discovered, again, for the directed and unweighted social network. NWB extracted from 46 to 209 groups for the timeframe (average 88 per timeframe, Table 7.10). Algorithm by Blondel et al. works very fast, time required to extract single time window on computer 1. from Table 7.1 was below 30 seconds. All timeframes were extracted in less than 7 minutes. The average size of the group this time was 64 nodes (Table 7.10), the smallest groups had size 2, and the biggest one was 750 in time window 8.

| **Time window** | **No of nodes in network** | **Number of groups** | **Avg size of a group** |
|---|---|---|---|
| 1 | 4479 | 55 | 82,9 |
| 2 | 4502 | 82 | 55,6 |
| 3 | 4382 | 95 | 46,6 |
| 4 | 4456 | 117 | 38,4 |
| 5 | 4527 | 103 | 44,4 |
| 6 | 4856 | 72 | 68,4 |
| 7 | 4893 | 64 | 77,7 |
| 8 | 4950 | 64 | 78,6 |
| 9 | 4965 | 47 | 108 |
| 10 | 4904 | 57 | 87,6 |
| 11 | 4870 | 61 | 81,2 |
| 12 | 4771 | 73 | 66,3 |
| 13 | 4670 | 139 | 33,8 |
| 14 | 4414 | 210 | 21,1 |
| **Avg** | **4689** | **89** | **63,6** |

Table 7.10. Results of Blondel et al. method extraction.

### *7.3.1. GED Method.*

As previously for data grouped with CPM method, the GED method have been run with different value of $\alpha$ and $\beta$ thresholds, the results are presented in Table 7.11. The time needed for single run was about 13 minutes. The thresholds for the *forming* and *dissolving* event was again set to 10%.

| Threshold | | Number of | | | | | | | |
|---|---|---|---|---|---|---|---|---|---|
| **α** | **β** | **form** | **dissolve** | **shrink** | **growth** | **continue** | **split** | **merge** | **total** |
| 50 | 50 | 39 | 23 | 187 | 167 | 135 | 411 | 269 | 1231 |
| 50 | 60 | 39 | 23 | 181 | 161 | 135 | 378 | 275 | 1192 |
| 50 | 70 | 39 | 23 | 179 | 156 | 135 | 338 | 280 | 1150 |
| 50 | 80 | 39 | 23 | 178 | 153 | 135 | 294 | 283 | 1105 |
| 50 | 90 | 39 | 23 | 164 | 143 | 134 | 250 | 293 | 1046 |
| 50 | 100 | 39 | 23 | 154 | 143 | 134 | 224 | 293 | 1010 |
| 60 | 50 | 39 | 23 | 181 | 166 | 135 | 417 | 237 | 1198 |
| 60 | 60 | 39 | 23 | 176 | 159 | 134 | 383 | 244 | 1158 |
| 60 | 70 | 39 | 23 | 174 | 155 | 134 | 338 | 247 | 1110 |
| 60 | 80 | 39 | 23 | 171 | 151 | 134 | 294 | 251 | 1063 |
| 60 | 90 | 39 | 23 | 156 | 140 | 133 | 250 | 262 | 1003 |
| 60 | 100 | 39 | 23 | 148 | 140 | 133 | 218 | 262 | 963 |
| 70 | 50 | 39 | 23 | 169 | 164 | 134 | 429 | 216 | 1174 |
| 70 | 60 | 39 | 23 | 163 | 158 | 131 | 396 | 219 | 1129 |
| 70 | 70 | 39 | 23 | 164 | 154 | 130 | 345 | 221 | 1076 |
| 70 | 80 | 39 | 23 | 159 | 150 | 130 | 299 | 225 | 1025 |
| 70 | 90 | 39 | 23 | 144 | 139 | 129 | 245 | 236 | 955 |
| 70 | 100 | 39 | 23 | 137 | 138 | 129 | 204 | 237 | 907 |
| 80 | 50 | 39 | 23 | 162 | 165 | 134 | 436 | 180 | 1139 |
| 80 | 60 | 39 | 23 | 157 | 158 | 130 | 402 | 178 | 1087 |
| 80 | 70 | 39 | 23 | 156 | 152 | 129 | 350 | 176 | 1025 |
| 80 | 80 | 39 | 23 | 151 | 147 | 127 | 304 | 177 | 968 |
| 80 | 90 | 39 | 23 | 138 | 140 | 126 | 235 | 184 | 885 |
| 80 | 100 | 39 | 23 | 128 | 140 | 126 | 191 | 184 | 831 |
| 90 | 50 | 39 | 23 | 157 | 172 | 133 | 442 | 126 | 1092 |
| 90 | 60 | 39 | 23 | 153 | 161 | 129 | 407 | 124 | 1036 |
| 90 | 70 | 39 | 23 | 152 | 152 | 128 | 355 | 118 | 967 |
| 90 | 80 | 39 | 23 | 146 | 139 | 126 | 310 | 116 | 899 |
| 90 | 90 | 39 | 23 | 133 | 130 | 121 | 228 | 114 | 788 |
| 90 | 100 | 39 | 23 | 116 | 131 | 121 | 178 | 113 | 721 |
| 100 | 50 | 39 | 23 | 160 | 168 | 133 | 439 | 106 | 1068 |
| 100 | 60 | 39 | 23 | 156 | 154 | 129 | 404 | 104 | 1009 |
| 100 | 70 | 39 | 23 | 155 | 144 | 128 | 352 | 97 | 938 |
| 100 | 80 | 39 | 23 | 149 | 129 | 126 | 307 | 95 | 868 |
| 100 | 90 | 39 | 23 | 133 | 110 | 121 | 228 | 83 | 737 |
| 100 | 100 | 39 | 23 | 114 | 109 | 120 | 178 | 80 | 663 |

Table 7.11. The results of GED computation on disjoint groups extracted by Blondel et al. method.

The total number of events found with thresholds equal 50% was 1231, and with thresholds equal 100% only 663. This indicates that parameters $\alpha$ and $\beta$ influence on number of events found even more than in case of CPM method. The linear relationship between increase of $\alpha$ threshold and reduction in number of merging and growing events is preserved, Figure 7.7 and Figure 7.8.

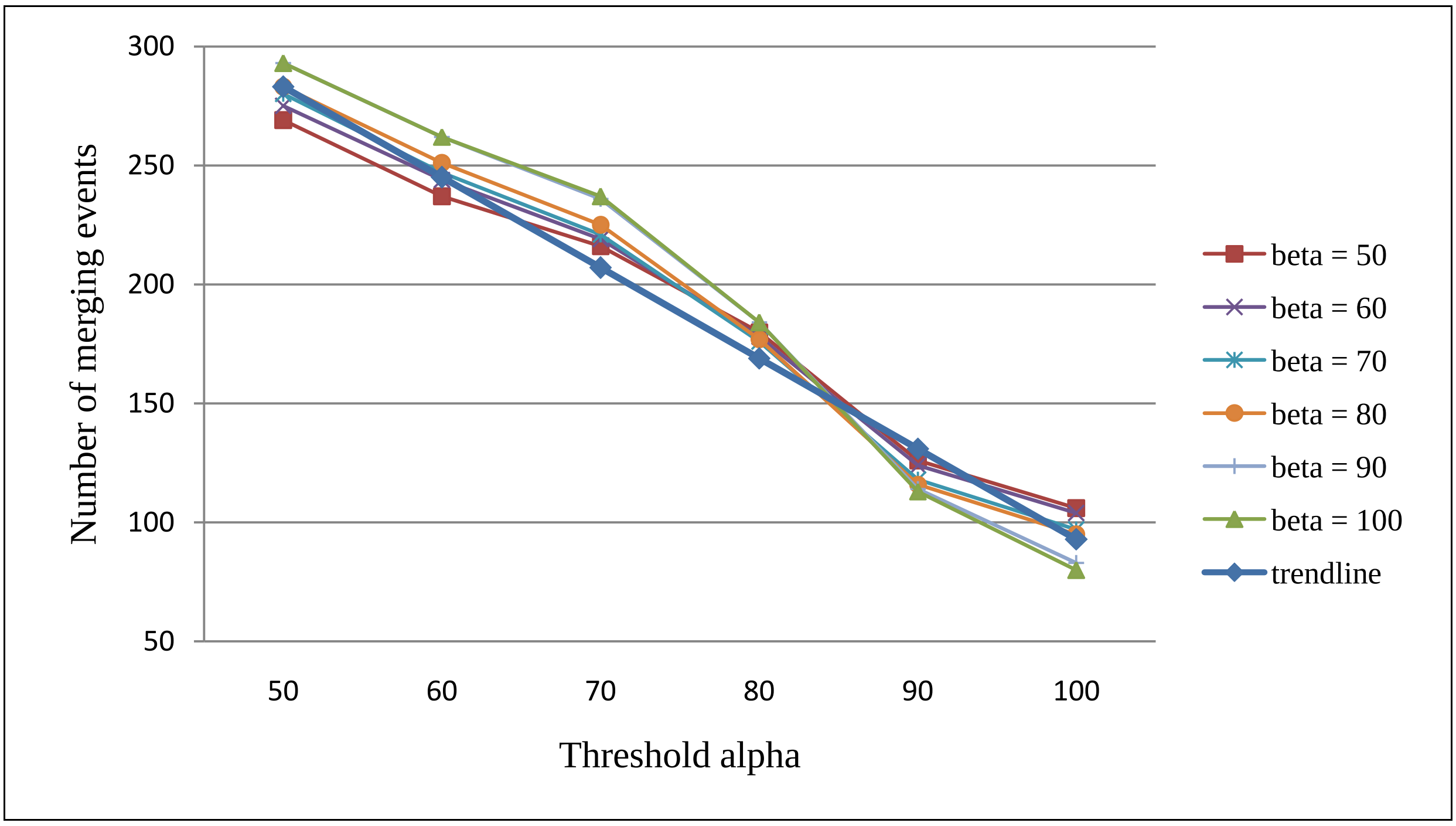

Figure 7.7. Number of merging events for different values of alpha and beta thresholds.

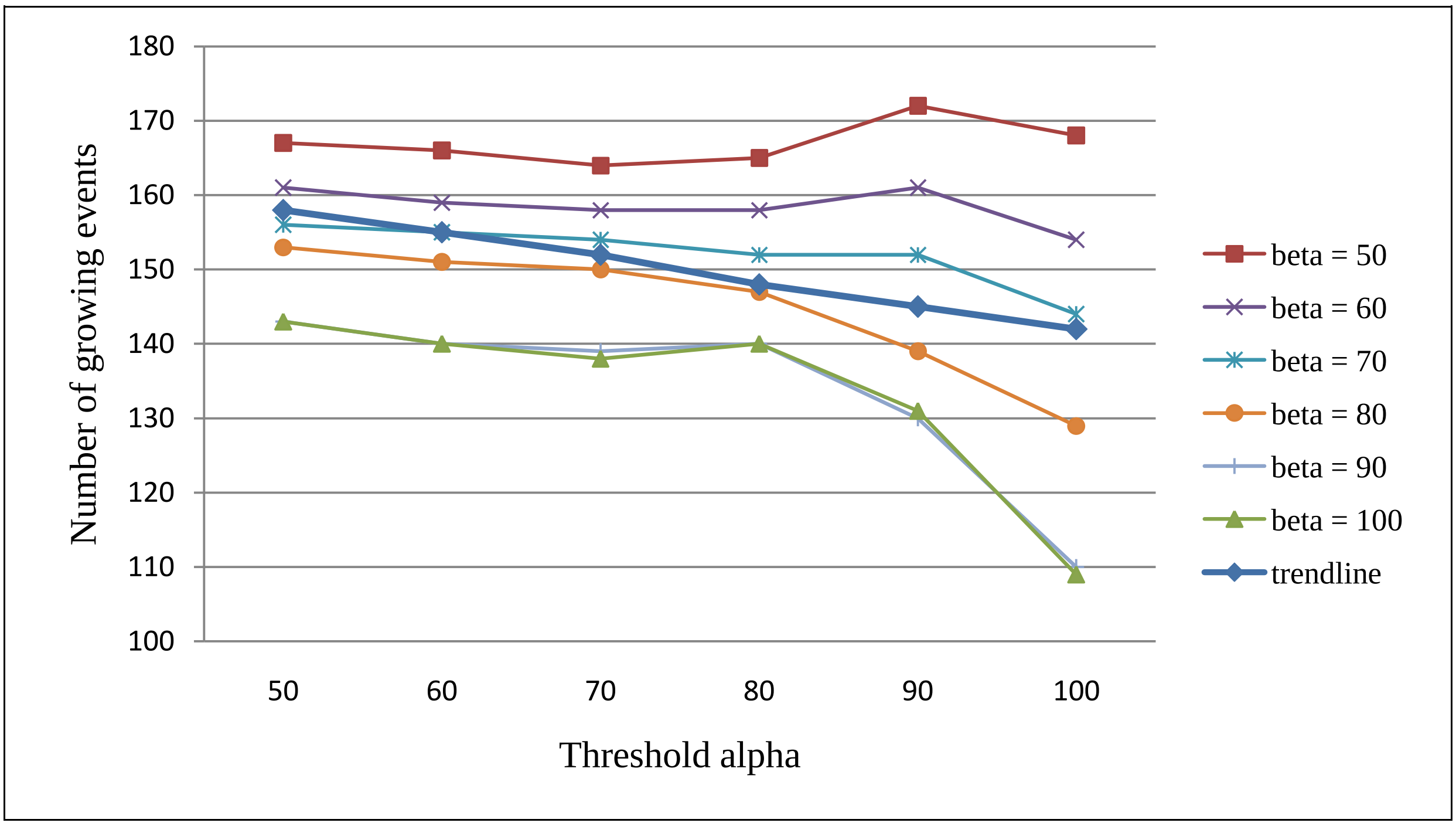

Figure 7.8. Number of growing events for different values of alpha and beta thresholds.

The linear relationship between increase of $\beta$ threshold and reduction in number of splitting and shrinking events is also preserved, Figure 7.9 and Figure 7.10.

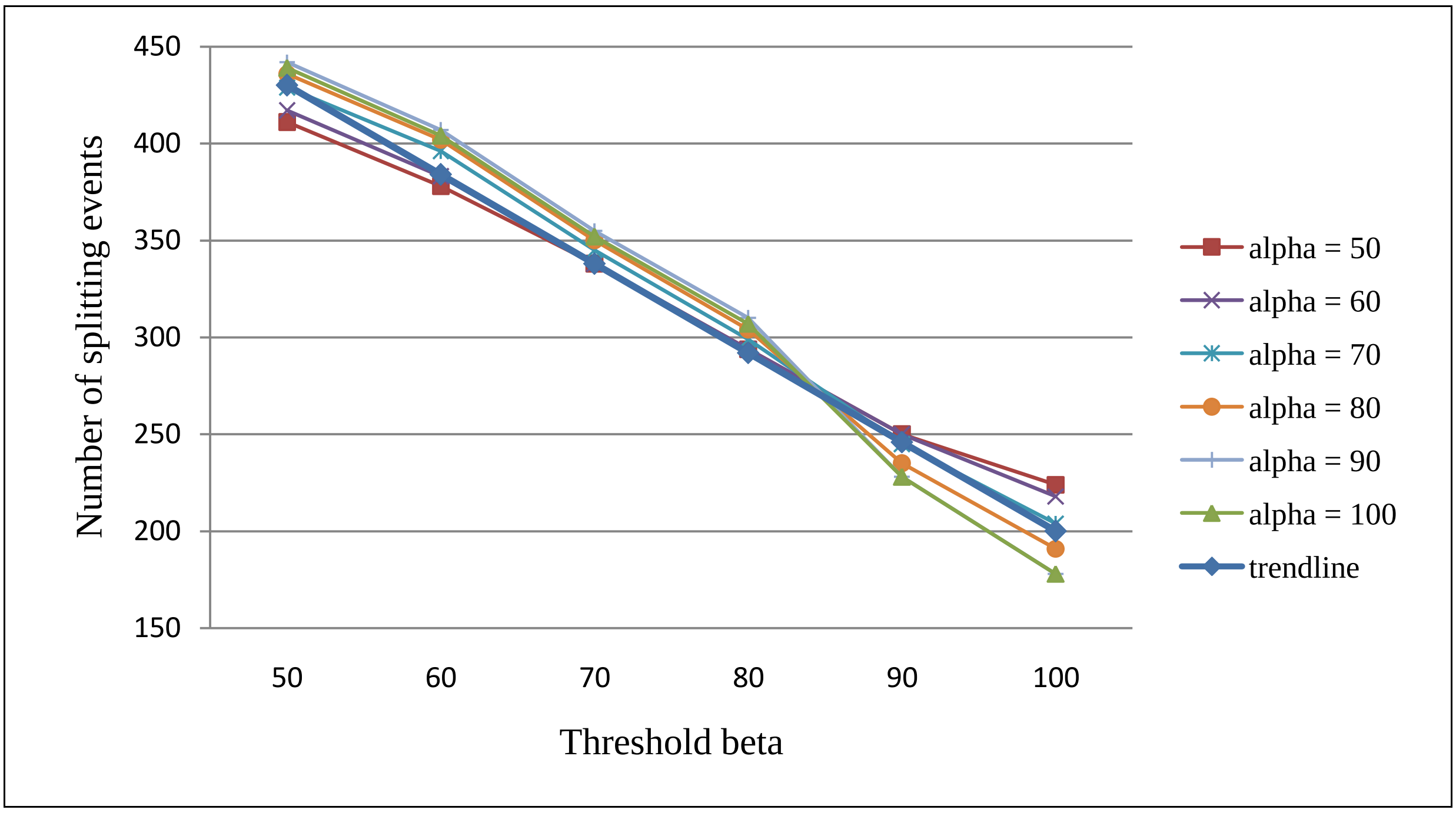

Figure 7.9. Number of splitting events for different values of beta and alpha thresholds.

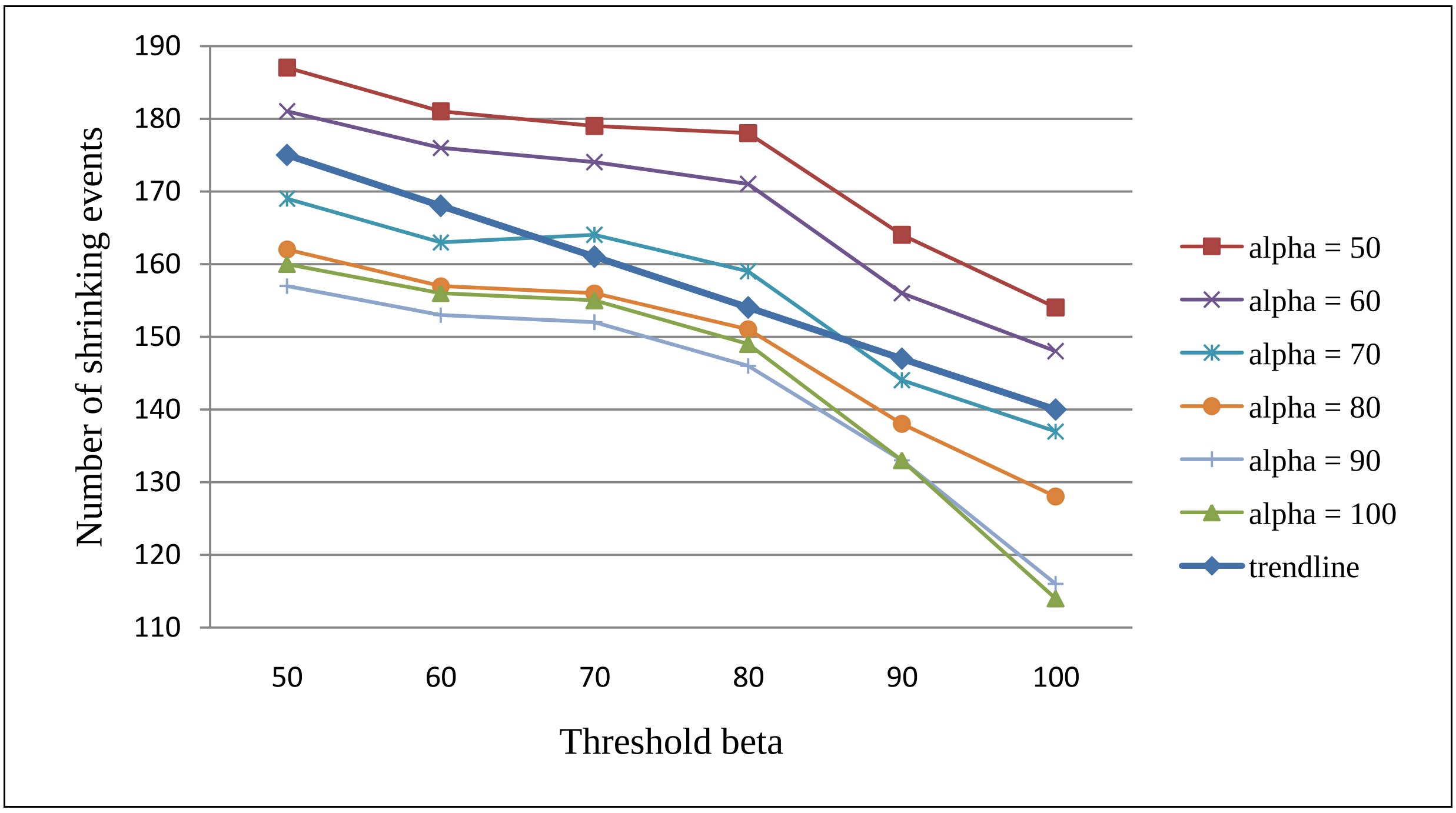

Figure 7.10. Number of shrinking events for different values of beta and alpha thresholds.

Opposed to the experiment with CPM method, this time dissolving event occurred less times than forming event. The reason is design of method by Blondel et al., which detects groups consisting even of only two members. Therefore it is hard for group to vanish completely. Unless the threshold is set above 50%, slightly changes by few percent will not generate more forming/dissolving events. GED method found 120 continue events when both inclusions of groups are equal to 100%, which corresponds to *continuation* event in Asur et al. method.

The experiment confirmed that GED method can be successfully used for both, overlapping and disjoint groups. If one needs overlapping groups for a small network then

CPM can be used, but if one needs to extract groups very fast and for a big network then the method proposed by Blondel et al. can be utilized. That is a big advantage because most methods can be used only for either overlapping or disjoint groups.

#### *7.3.2. Method by Asur et al.*

The method provided by Asur et al. needed almost 6 hours to calculate events between groups in all fourteen time windows. The overlapping threshold for merging and splitting events was again set to 50%. The total number of events found by Asur et al. method is 747, from which 120 are continuation, 23 are forming, 16 are dissolving, 255 are merging and 333 are splitting.

As previously, small number of continuing, forming, and dissolving events is caused by too rigorous conditions. In turn, great number of merging (splitting) events is a result of low overlapping threshold for merge (split) and lack of growing (shrinking) event.

In contrast to previous experiment, the number of events found on data grouped by Blondel et al. method is not overestimated. This time method by Asur et al. has not generate any anomalies, what confirms the assumption that the method is designed for disjoint groups.

All other outcomes from experiment conducted on overlapping groups were confirmed in experiment with disjoint groups.

#### *7.3.3. GED Method vs. Asur et al. Method*

GED method needed less than 8 hours to calculate whole Table 7.11, while one run of Asur et al. method was almost 6 hours. Single run of GED method was only 13 minutes, so it is still much faster than method by Asur et al.

The GED method run with thresholds equals 50% found 613 events which method by Asur et al. has not. Again, big lack in results obtained with Asur et al. method is caused mostly by rigorous conditions for assigning events and almost no flexibility of the method. Like in case of CPM method, Asur et al. method found events which GED method skipped because of thresholds value. Reducing the thresholds effected in not omitting mentioned groups.

Provided considerations confirms that GED method is better than Asur et al. method for overlapping as well as for disjoint methods of grouping social network.

### 7.4. Experiment Based on Different User Importance Measures

In the last experiment GED method was run: (1) with degree centrality measure instead of social position measure and (2) without any measure, in order to investigate influence of the measure on calculations of inclusion values and also on results of the method. Like in case of the first experiment, overlapping groups extracted with CPM were used.

The results obtained with degree centrality as a measure of user importance and results derived without any measure are very similar to the results obtained with social position measure, Table 7.12.

| **Measure** | **Execution time [min]** | **Events found** | **Threshold** | |
|---|---|---|---|---|
| | | | $\alpha$ | $\beta$ |
| Social Position | 6 | 1470 | 70 | 70 |
| Degree Centrality | 5:55 | 1447 | 70 | 70 |
| No measure | 5:30 | 1483 | 70 | 70 |

Table 7.12. Results of GED method with different user importance measures.

Execution time for GED with degree centrality was slightly better than for GED with social position because degree centrality value is given as a integer, while the type for social position value is float. Summing integers is faster than summing floats, thus the difference. Of course the best execution time was for GED without a measure as a effect of less calculations needed to proceed. Although, the number of events found in all three cases is more or less the same, it can be observed that GED without user importance measure found more events than GED with any of the measures. It is a consequence of the inclusion formula (see equation 4.2) which consists of two fractions The second fraction is always present, whether GED is run with or without user importance measure, but the first one occurs only when a measure is used. Therefore when calculating inclusions of two groups with a measure, it is almost always lower than without any measure. The exceptions are groups where inclusions are equal 100% and groups which do not share any nodes (inclusions equals 0%). And here comes the question again: why GED uses a measure of user importance, since it is obvious that it will lower the inclusion? The answer already provided in Section 4.2 this time is supported by clear evidences.

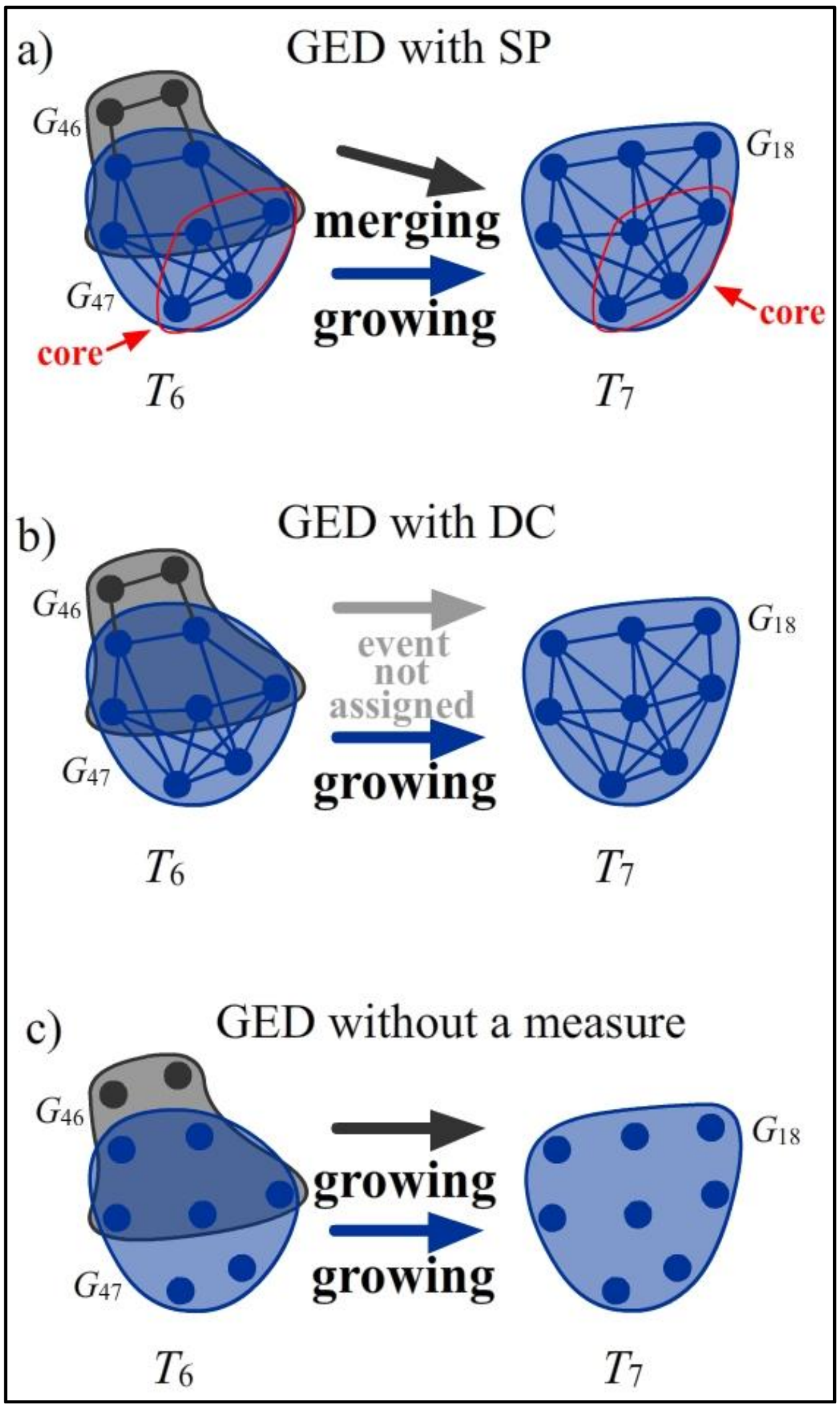

Figure 7.11. Events assigned by GED method with different user importance measures. a) GED with social position measure, red colour marks the core of the group b) GED with degree centrality c) GED without a measure.

As illustrated in the example in Figure 7.11, two communities $G_{46}$ and $G_{47}$ from timeframe $T_6$ overlaps by five members and both groups have the same size – seven members. In the next timeframe $T_7$ there is only one group $G_{18}$ which consists of all members from the group $G_{47}$ from the previous timeframe, and one new member. Two members from the community $G_{46}$ vanished in the following time window.

GED method run with social position measure assigned growing event to the community $G_{47}$ and merging event to the group $G_{46}$. GED method run with degree centrality measure also assigned merging event to the group $G_{47}$, but did not assign any event to the community $G_{46}$. Finally, GED without any user importance measure assigned growing events to both groups from timeframe $T_6$.

To have a closer look into the first case, social position of members is presented in Table 7.13. It is clearly visible that the core of the blue group from timeframe $T_6$ is identical to the core of the blue group from the next time window $T_7$. The situation is marked with red colour in the Figure 7.11a and with red dots in the Table 7.13. Additionally, members occurring in all the groups are marked green. Now it is obvious that GED with social position measure assigned growing event to group $G_{47}$ because it is almost identical to group $G_{18}$, and “only” merging event to group $G_{46}$ because the cores of both groups have nothing in common. It has to be mentioned once again that, thanks to the user importance measure, GED method takes into account both the quantity and quality of the group members providing very accurate and specified results.

| **Group** | **Time window** | **Node** | **SP** | **Rank** |
|---|---|---|---|---|
| 46 | 6 | 1443 | 1,48 | 1 |
| 46 | 6 | 3145 | 1,33 | 2 |
| 46 | 6 | 7564 | 0,96 | 3 |
| 46 | 6 | 1326 | 0,86 | 4 |
| 46 | 6 | 11999 | 0,85 | 5 |
| 46 | 6 | 14151 | 0,77 | 6 |
| 46 | 6 | 621 | 0,75 | 7 |
| 47 | 6 | 2066• | 1,31 | 1 |
| 47 | 6 | 7328• | 1,30 | 2 |
| 47 | 6 | 7564• | 1,28 | 3 |
| 47 | 6 | 11999• | 1,04 | 4 |
| 47 | 6 | 1326 | 0,80 | 5 |
| 47 | 6 | 14151 | 0,67 | 6 |
| 47 | 6 | 621 | 0,60 | 7 |
| 18 | 7 | 2066• | 1,49 | 1 |
| 18 | 7 | 7328• | 1,35 | 2 |
| 18 | 7 | 7564• | 1,29 | 3 |
| 18 | 7 | 11999• | 1,24 | 4 |
| 18 | 7 | 1326 | 0,75 | 5 |
| 18 | 7 | 14151 | 0,71 | 6 |
| 18 | 7 | 621 | 0,66 | 7 |
| 18 | 7 | 4632 | 0,51 | 8 |

Table 7.13. Social position of members presented in Figure 7.11a.

GED method run with degree centrality measure was even more strict in the studied case, Figure 7.11b. Low degree centrality within the group $G_{46}$ causes that no event was assigned. In turn, similar structure between groups $G_{47}$ and $G_{18}$ effects in assigning merging event. Structure of all groups and degree centrality of all members is presented in Table 7.14. Again, green colour marks members occurring in all groups.

| Group | Time window | Node | DC | Rank |
|---|---|---|---|---|
| 46 | 6 | 11999 | 3 | 1 |
| 46 | 6 | 14151 | 3 | 1 |
| 46 | 6 | 1443 | 2 | 3 |
| 46 | 6 | 3145 | 2 | 3 |
| 46 | 6 | 7564 | 2 | 3 |
| 46 | 6 | 1326 | 2 | 3 |
| 46 | 6 | 621 | 2 | 3 |
| 47 | 6 | 2066 | 5 | 1 |
| 47 | 6 | 7328 | 5 | 1 |
| 47 | 6 | 7564 | 4 | 3 |
| 47 | 6 | 11999 | 4 | 3 |
| 47 | 6 | 1326 | 4 | 3 |
| 47 | 6 | 14151 | 3 | 6 |
| 47 | 6 | 621 | 3 | 6 |
| 18 | 7 | 7564 | 7 | 1 |
| 18 | 7 | 7328 | 5 | 2 |
| 18 | 7 | 2066 | 5 | 2 |
| 18 | 7 | 11999 | 5 | 2 |
| 18 | 7 | 1326 | 5 | 2 |
| 18 | 7 | 14151 | 4 | 6 |
| 18 | 7 | 621 | 4 | 6 |
| 18 | 7 | 4632 | 3 | 8 |

Table 7.14. Degree centrality of members presented in Figure 7.11a.

Figure 7.11c expresses in the best way how GED method without a user importance measure understands the communities. There is no core, all members are equal and relations between them are not considered at all. Such simplification causes that events assigned to the groups are not the most adequate to situation (but only when comparing with events assigned by GED with user importance measure). Having only information about members in the groups, not about their relations, events are assigned correctly. So, if researchers investigating group evolution are not interested in groups structure and relations between members, a simpler and faster version of GED method may be successfully used. However, if there is enough time and possibility to calculate any user importance measure, it is recommended to use GED method in the original version.

## 8. Conclusions and Future Work

The number of social systems in which people are communicating with each other is rising at an unprecedented rate. That creates endless need to analyse them. One part of such analysis is communities detection and investigation of their evolution over time in order to understand the mechanisms governing the development and variability of social groups.

The research in area of extracting social groups showed that dozens of methods are existing, and depending on the characteristic of the social network one or several of them can be successfully used to detect groups. As Section 7.3 shows, methods based on fast modularity are much faster than methods based on cliques for detecting groups from large datasets. The problem was formulated in the research question no. 1.

Thereby, groups are prepared for further analysis in which their tendencies and behaviour may be studied. To do so, proper methods or techniques are required. The research in this area revealed gap in the knowledge since the existing methods are either computational too expensive or too less accurate or simply not able to explore basic types of social networks such as overlapping networks and disjoint networks. To meet most demanding needs of researchers facing problem of tracking changes in community life the new method called GED (Group Evolution Discovery) was proposed and evaluated in this thesis. Answering the research question no 2. whole process of tracking evolution is described in Section 4.3. Based on the information about social network at succeeding intervals of time, especially based on the social position of members within communities extracted from these networks method assigns events to the groups indicating changes. A set of the events assigned to a single community throughout several timeframes represents the history of evolution of this group. Such history allows to study birth (forming), death (dissolving), splitting and merging, shrinking and growing, and finally stagnation (continuing) of the community, which are the most common event types occurring in social group evolution – the research question no. 3.

The requirements for the new method were substantial, the method had to:

- be accurate – catch all possible evolutions of a group,
- be flexible – allows to adjust method to the ones needs,
- work fast and with low computational costs,
- be easily in implementation,
- be intuitive.

The results of experiments and comparison with the existing methods for discovering changes in community life (Section 7) leads to the conclusion that desired features were achieved exemplary, and the new method may become one of the best method for tracking group evolution.

In order to answer research question no. 4, two experiments based on both types of groups were conducted and Sections 7.2 and 7.3 presents results obtained with various methods for tracking group evolution. More specifically, GED method is incredibly faster and flexible than methods by Palla et al. and by Asur et al. Additionally, GED method is more accurate than method by Asur et al. and much more specific when it comes to naming changes of a group than method by Palla et al.

Experiments conducted on Asur et al. method (Section 7.2.2 and Section 7.3.2) showed that this method is barely flexible and therefore skips evolutions, what disqualifies the method for any studies regarding group evolutions. Furthermore, the method generates anomalies, when working on overlapping groups, which implies that method is designed for working only on disjoint groups – the research question no. 5. Other investigated aspects: computation cost, ease of implementation and execution time, came out promising, however in comparison with GED method they were much worst.

Experiment run with Palla et al. method (Section 7.2.4) demonstrated that great idea and unique design of the method are wasted by the lack of formula for assigning events and by few smaller ambiguities in the method. Despite the fact that method by Palla et al. is computationally extremely expensive and can be use only on groups extracted by CPM, the method is still considered the best algorithm tracking evolution for overlapping groups – the research question no. 5. Although, straight comparison with GED method revealed that method by Palla et al. is very cumbersome and useful only to test accuracy of GED method.

The last experiment demonstrates usefulness of the social position measure and convinces that GED method should be always run with user importance measure in order to fully benefit all features of the algorithm.

Summing up, the GED method was designed to be as much flexible as possible and fitted to both, overlapping and disjoint groups but also to have low and adjustable computational complexity. The GED method, described and analysed in this paper, uses not only the group size and comparison of groups members, but also takes into account their position and importance in the group to determine what happened with the group.

The first results of GED method were presented in [70] and [71]. Next, more exhaustive analysis was conducted within this thesis, but the method is still in the development phase. In the next step method will be extended by migration event, which focuses on the core of the group (the most important members – leaders) and tracks whether members outside the core followed the leaders or not.

## Acknowledgement

This work was supported by MSc. Piotr Bródka, Ph.D. Przemysław Kazienko – Wrocław University of Technology, Institute of Informatics and Social Network Group – Wrocław University of Technology.